**On the Dynamical and Thermodynamic Constraints of Axisymmetric Tropical Cyclones under Non-Symmetric-Neutrality**

Chau-Lam Yu[1,2]

[1]Atmospheric Sciences Research Center, SUNY University at Albany, Albany, NY, USA

[2]Department of Atmospheric and Environmental Sciences, SUNY University at Albany, Albany, NY, USA

*Corresponding author address*: Chau-Lam Yu, New York State Mesonet, ETEC Building - Harriman Campus, 1200 Washington Avenue, Room 360E, Albany, NY 12226

E-mail: cyu7@albany.edu

**Abstract**

The potential intensity (PI) theory of tropical cyclones (TCs) provides a reasonable estimate of the steady-state intensity in a quiescent environment. The theory relies on the symmetric neutrality (SN) assumption, where absolute angular momentum ($M$) surfaces are parallel to the saturation entropy ($s^*$) surfaces within the eyewall above the boundary layer. However, existing theories do not explain how these variables constrain the vortex structure and maximum tangential wind ($v_{max}$) under non-symmetric neutrality (non-SN) conditions.

This study relaxes the SN assumption to derive a generalized $v_{max}$ formula that summarizes the dynamical and thermodynamic constraints on the vortex structure and intensity under non-SN conditions. It is proven that under non-SN conditions, the gradient of $s^*$ with respect to $M$ holding temperature ($T$) constant constrains the curvature of the $M$ surfaces throughout the saturated eyewall, thereby determining the balanced intensity above the TC boundary layer. In addition, generalized expressions for the unbalanced and frictional contributions to $v_{max}$ are also derived.

Verifying against axisymmetric simulations, this generalized formula accurately quantifies the balanced, unbalanced, and frictional contributions during the rapid intensification (RI). Despite being diagnostic, it offers valuable insights into TC intensification under non-SN conditions: (1) before reaching SN, the $s^*$ gradient should be computed holding $T$ fixed to quantify the balanced wind component. (2) During TC intensification, the $s^*$ gradient distinctly increases with height along the eyewall updraft, confirming that SN assumption is not valid during RI. The implications of these findings on the vortex structure and the upper-tropospheric mixing during RI are examined.

## 1. Introduction

As the long-lasting eyewall convection of a tropical cyclone (TC) transports high-entropy air into the upper troposphere, convective and symmetric instabilities are released and eventually neutralized, resulting in a steady-state vortex structure with zero moist potential vorticity (Emanuel 1986, E86 hereafter). This steady-state vortex structure is symmetrically neutral, where the isolines of saturation entropy $s^*$ and absolute angular momentum $M$ are congruent to streamfunction $\psi$ in the free troposphere. The symmetric neutrality (SN) assumption, along with gradient wind and hydrostatic balance, has long been known to constrain the steady-state structure and intensity of tropical cyclones (E86; Emanuel and Rotunno 2011, ER11 hereafter; D. K. Lilly 1979, 1986 (unpublished manuscripts); Tao et al. 2020). In addition, the SN condition is the core assumption of various theories of the impacts of dissipative heating (Bister and Emanuel 1997), intrusion of low entropy air into TC inner core, i.e., ventilation (Tang and Emanuel 2010; 2012a,b), the potential size of TCs (Wang et al. 2022), etc. Building upon the SN and steady-state assumptions, Bryan and Rotunno (2009a,b, BR09a,b) derived a quantitative estimate for the contribution of the unbalanced processes on the maximum tangential wind at and above the boundary layer top.

In the past decade, research efforts have gradually shifted from steady-state centric to focusing more on TC intensification. Emanuel (2012, E12) demonstrated that assuming congruence between $s^*$ and $M$ surfaces in the free troposphere, while allowing for crossing between these surfaces and streamfunction isolines, permits the temporal evolution of intensity, yielding a time-dependent model that captures the later intensification period. A similar time-dependent model was later derived using thermodynamic cycle (Ozawa and Shimokawa 2015), even though SN was assumed implicitly[1]. Despite this advancement, the E12 time-dependent model has clear limitations, as it does not accurately capture the early intensification evolution and requires a subjectively determined "ignition time" to match the analytic solution with numerical model simulations. Consequently, it cannot be used to predict the onset timing of intensification.

These shortcomings of the E12 model exposed the fundamental limitations of the SN assumption. As confirmed by other studies (e.g., Peng et al. 2018, 2019), the SN assumption is not valid until the later stages of the intensification period. Recent developments in time-dependent models have attempted to incorporate the effects of non-SN by introducing several ad-hoc parameters (Wang et al. 2021a,b; Li et al. 2024). While those empirically determined parameters help to identify the dependence of intensification rate that matches the numerical simulations, it is difficult to guarantee that the resulting dependence is universally true and not specific only to the set of numerical experiments. More importantly, it is challenging to discern the underlying physics that determines specific values of the parameters.

[1] The cyclic return of parcels to initial state is true only at steady state. Since $M$ and $s^*$ are approximately conserved in free troposphere, the steady state requirement thus forces both variables to be congruent with $\psi$. Hence, SN is assumed implicitly.

This recent development motivates a more thorough examination of the axisymmetric moist vortex dynamics before the SN assumption becomes valid. Many questions surrounding the dynamics of non-symmetrically neutral (non-SN) vortices remain unanswered. For instance, the PI theory showed that the balanced boundary layer top tangential wind is dynamically linked to the total derivative of $s^*$ w.r.t $M$, i.e., $ds^*/dM$, and the temperature difference between the boundary layer top and the outflow level. However, the term $ds^*/dM$ is mathematically well-defined only when $s^*$ and $M$ isolines are congruent. When the SN condition is not met, the physical quantity $\frac{\partial s^*}{\partial M}$ is not mathematically defined, because the direction of evaluation is not explicitly specified. This ambiguity raises the important question: among all possible directions to compute $\frac{\partial s^*}{\partial M}$, *which direction is consistent with balanced dynamics and physically related to the balanced tangential wind maximum at the low level?* In the recent time-dependent TC intensification theory, Wang et al. (2021b) and Li et al. (2024) computed $\frac{\partial s^*}{\partial M}$ along the direction perpendicular to $s^*$ contour, i.e., $\left(\frac{\partial s^*}{\partial M}\right)_{\perp s^*}$ with $\perp s^*$ denoting the normal direction of $s^*$ isoline. However, no physical justification was provided for their choice of evaluation direction of $\frac{\partial s^*}{\partial M}$ (Li et al. 2024; and personal communication). To address these questions, we will revisit the derivation of the PI theory without making the SN assumption to uncover how the established results from previous studies (E86; BR09b) can be extended into non-SN regime.

The primary objective of the work is to establish a fundamental framework that captures the essential moist vortex dynamics for non-SN vortices, thereby providing a rigorous foundation for studying the TC intensification process and for the future development of time-dependent theory. Using the radial and vertical momentum equations and the modified first law of thermodynamics, several fundamental results are obtained without making the SN assumption, including a new energetic property that constrains the structure of a balanced vortex, and a generalized (diagnostic) tangential wind formula that encompasses both the balanced and unbalanced components under non-SN condition. It should be acknowledged that asymmetric processes are an intrinsic component of the TC intensification even in the absence of vertical shear and vortex translation (Persing et al. 2013). We choose to focus on the axisymmetric modeling framework since it, to a large extent, can capture the major component of the TC intensification and can provide important insight into underlying dynamics.

The remainder of this paper is organized as follows. We derive the theoretical findings in section 2. Section 3 describes the numerical experiments and the evaluation methods. Section 4 presents a detailed verification of the theoretical results, while section 5 discusses the implications of these findings. Section 6 summarizes the major findings and limitations of this work.

## 2. Theory

### *a. The balanced component*

For direct comparison with the results from E86 and ER11, we first consider the balanced case in gradient wind and hydrostatic balance. The fully generalized form without balanced assumptions will be presented in section 2b. In the radius-pressure $(r, p)$ coordinate system and assuming axisymmetry, the gradient wind and hydrostatic balance equations may be written as

$$\frac{\partial \phi}{\partial r} = \frac{M^2}{r^3} - \frac{1}{4} f^2 r \, , \tag{1}$$

$$\frac{\partial \phi}{\partial p} = -\alpha_d \, , \tag{2}$$

where $M = rv + \frac{1}{2} f r^2$ the absolute angular momentum, $\phi$ the geopotential; $\alpha_d = 1/\rho_d$ the dry air specific volume. As discussed in BR09b, the above formulation neglects the contributions of moisture and hydrometeors to the value of density $\rho$ of moist air. The thermal wind can be obtained by cross-differentiating (1) with respect to $p$ and (2) with respect to $r$ and taking their difference. The result is

$$-\frac{1}{r^3} \left( \frac{\partial M^2}{\partial p} \right)_r = \left( \frac{\partial \alpha_d}{\partial r} \right)_p . \tag{3}$$

Similar to E86, we assume the air within the eyewall updraft is saturated and define saturation entropy $s^* = c_p \ln T - R_d \ln p_d + \frac{L_v q^*}{T}$, where $p_d$ is dry air density, $T$ the temperature, $q^*$ the saturation vapor mixing ratio, $c_p$ the heat capacity at constant pressure, $R_d$ the ideal gas constant of dry air, and $L_v$ the latent heat of vaporization. This entropy definition satisfies the modified first law of thermodynamics $Tds^* = c_p dT - \alpha_d dp + L_v dq^*$, where $p$ is the pressure of moist air. Writing $\left(\frac{\partial \alpha_d}{\partial r}\right)_p$ as $\left(\frac{\partial \alpha_d}{\partial s^*}\right)_p \left(\frac{\partial s^*}{\partial r}\right)_p$ and combining with Maxwell relation $\left(\frac{\partial \alpha_d}{\partial s^*}\right)_p = \left(\frac{\partial T}{\partial p}\right)_{s^*}$ (BR09b) and dividing both sides of equation (3) with $\left(\frac{\partial M}{\partial r}\right)_p$ yields

$$\frac{2M}{r^3}\left(\frac{\partial r}{\partial p}\right)_M = \left(\frac{\partial T}{\partial p}\right)_{s^*}\left(\frac{\partial s^*}{\partial M}\right)_p \ . \tag{4}$$

E86 and many subsequent studies proceeded by making the SN assumption, which is strictly true only when the TC evolves into a steady-state. But in this study, we will show that simplification of equation (4) can be achieved without making the SN assumption. As shown in appendix A, the term $\left(\frac{\partial T}{\partial p}\right)_{s^*}\left(\frac{\partial s^*}{\partial M}\right)_p$ can be written as $\left(\frac{\partial T}{\partial p}\right)_M\left(\frac{\partial s^*}{\partial M}\right)_p - \left(\frac{\partial T}{\partial M}\right)_p\left(\frac{\partial s^*}{\partial p}\right)_M = \det\left(\begin{bmatrix} \frac{\partial s^*}{\partial M} & \frac{\partial s^*}{\partial p} \\ \frac{\partial T}{\partial M} & \frac{\partial T}{\partial p} \end{bmatrix}\right)$, which is a Jocobian matrix determinant for transformation from $(M, p)$ space to $(s^*, T)$ space, denoted as $J\left(\frac{s^*,T}{M,p}\right)$. Based on the theorem of coordinate transformation, $\left(\frac{\partial T}{\partial p}\right)_{s^*}\left(\frac{\partial s^*}{\partial M}\right)_p dMdp = ds^*dT$ where $dMdp$ and $ds^*dT$ are the differential elements in $(M, p)$ and $(T, s^*)$ spaces. Therefore, by multiplying both sides of (4) with the differential element $dMdp$ in $(M, p)$ space, it can be simplified as

$$\frac{2M}{r^3}\left(\frac{\partial r}{\partial p}\right)_M dMdp = \left(\frac{\partial T}{\partial p}\right)_{s^*}\left(\frac{\partial s^*}{\partial M}\right)_p dMdp \ ,$$

$$dM^2 d\left(-\frac{1}{2r^2}\right) = ds^*dT \ . \tag{5}$$

Equation (5) is essentially the integral form of (4). Mathematically, it states that for any region of a balanced vortex, the corresponding areas in $\left(-\frac{1}{2r^2}, M^2\right)$ and $(s^*, T)$ spaces must be the same

$$\iint_D dM^2 d\left(-\frac{1}{2r^2}\right) = \iint_D dTds^* \ , \tag{6}$$

where $D$ is an arbitrary saturated region within a balanced vortex in physical space. Because equations (3) describes the balance between the azimuthal torques of centrifugal force and baroclinic torque[2] (both of which act to accelerate or decelerate the secondary circulation of the TC), equations (5) and (6) are thus the integral representation of the torque balance between these two forces within an arbitrary region $D$ of a balanced vortex, as demonstrated more concretely in Appendix C. In addition, using the Stoke's theorem, we show in supplementary material (S1) that equation (6) can also be interpreted as the balance of the net (virtual) work done by the nonconservative component of centrifugal force ($\oint_{\partial D} \frac{1}{2r^2} \nabla M^2 \cdot \boldsymbol{dl}$) and pressure gradient force ($\oint_{\partial D} s^* \nabla T \cdot \boldsymbol{dl}$) along an arbitrary closed boundary $\partial D$.

Direct verification of (5) and (6) requires simultaneous transformations of physical fields from $(r, p)$ physical space to $(M^2, -1/2r^2)$ and $(s^*, T)$ spaces, which can be challenging due to the complex structure of $M$ and $s^*$ fields during intensification. However, we note that (6) is also valid for a dry balanced vortex by replacing $s^*$ with dry entropy $s$. Figure 1 shows a verification example for a dry balanced vortex. By arbitrarily selecting two pairs of $M$ ($1\times10^6$ and $2.55\times10^6$ $m^2s^{-1}$) and $s$ (2475 and 2530 J $kg^{-1}$ $K^{-1}$) isolines, which define an arbitrary region $D$ (Fig. 1a), we confirm that the corresponding regions in $(-1/2r^2, M^2)$ space (Fig. 1b) and $(s, T)$ space (Fig. 1c) both have areas of about 211.3 [J $kg^{-1}$], thus verifying equations (5) and (6).

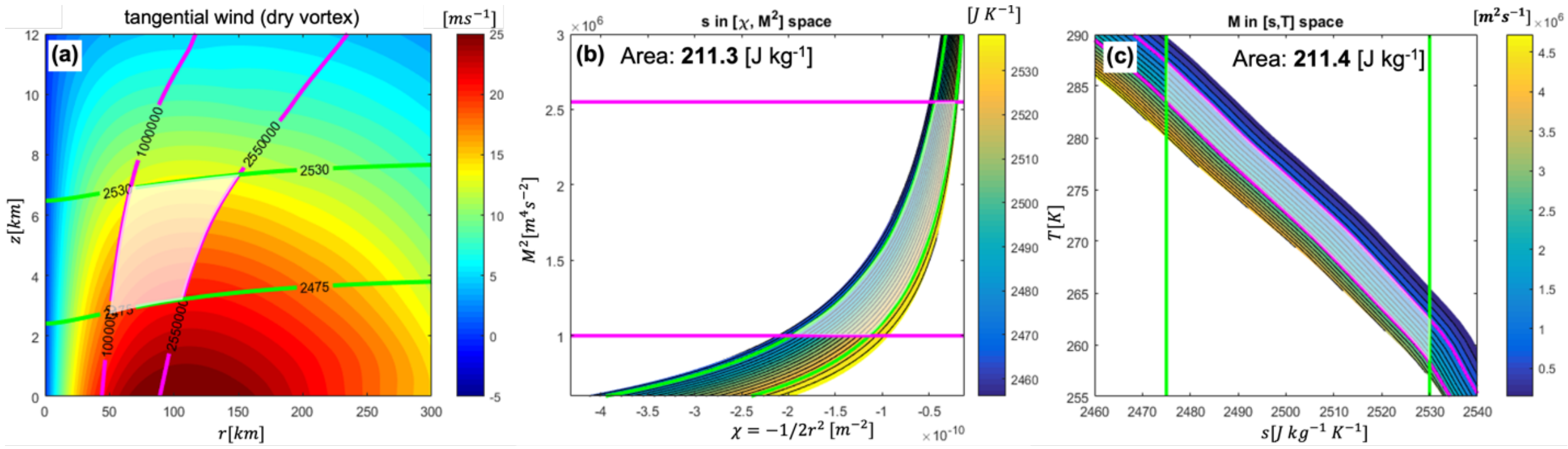

**Fig. 1:** (a) The tangential wind of a balanced, axisymmetric dry vortex. Magenta lines show the two $M$ surfaces ($10^6$ and $2.55 \times 10^6$ $m^2 s^{-1}$), while green lines show the two $s$ surfaces (2475 and 2530 J $kg^{-1}$ $K^{-1}$). The whitened portion is the region of interest ($D$). (b) Dry entropy $s$ in $(-1/2r^2, M^2)$ space. The same region of interest $D$ (whitened) has an area of 211.3 [J $kg^{-1}$] in this space. (c) is similar to (b), but for the $M$ in $(s, T)$ space, where the region of interest $D$ has an area of 211.4 [J $kg^{-1}$].

[2] As shown in the Appendix C, $\frac{2M}{r^3}\frac{\partial M}{\partial z} = \nabla \times \left(\left(\frac{v^2}{r} + fv\right)\hat{r}\right) \cdot \hat{\lambda}$ is centrifugal torque; whereas the baroclinic torque, defined as $J\left(\frac{\alpha, p}{r, z}\right) = \nabla \times (-\alpha \nabla p) \cdot \hat{\lambda}$, can also be written as $\nabla T \times \nabla s^* \cdot \hat{\lambda} = J\left(\frac{s^*, T}{r, z}\right)$ using the modified first law of thermodynamics.

We will now show that, without relying on the SN assumption, the energetic balance property (5) constrains the structure of the $M$ surfaces and determines the balanced tangential wind maximum at the low level under a general non-SN condition at any given time. Using the Jacobian identity $\left(\frac{\partial T}{\partial p}\right)_{s^*}\left(\frac{\partial s^*}{\partial M}\right)_p = \left(\frac{\partial T}{\partial p}\right)_M\left(\frac{\partial s^*}{\partial M}\right)_T$ (as derived in Eqs. A.3a and A.3b), we may rewrite Eq. (4) as follows:

$$\frac{2M}{r^3}\left(\frac{\partial r}{\partial p}\right)_M = \left(\frac{\partial T}{\partial p}\right)_M\left(\frac{\partial s^*}{\partial M}\right)_T . \tag{7}$$

Dividing both sides of (7) by $\left(\frac{\partial T}{\partial p}\right)_M$ yields

$$\frac{2M}{r^3}\left(\frac{\partial r}{\partial T}\right)_M = \left(\frac{\partial s^*}{\partial M}\right)_T . \tag{8}$$

Equation (8) explicitly relates the inverse slope of an $M$ surface in $(r, T)$ space, i.e., $\left(\frac{\partial r}{\partial T}\right)_M$, to the slope of a temperature surface in $(M, s^*)$ space, i.e., $\left(\frac{\partial s^*}{\partial M}\right)_T$. This relationship is the generalization of how $\frac{ds^*}{dM}$ constrains the shape (slope) of an $M$ surface and balanced intensity under symmetric neutrality, derived in E86. The physical picture of (8) is that as the $M$ surfaces of a balanced vortex bend towards the TC inner core to spin up the balanced vortex, the change of tangential wind along the $M$ surfaces results in a differential centrifugal force (plus Coriolis), which constitute an azimuthal torque, as shown in Fig. 2a. To balance this centrifugal torque, a baroclinic torque in the opposite direction (Fig. 2b) must exist by developing a radial gradient in density. In $(M, s^*)$ space, this density gradient is manifested as horizontal variation in temperature, which determines the slope of the temperature isolines, i.e., $\left(\frac{\partial s^*}{\partial M}\right)_T$.

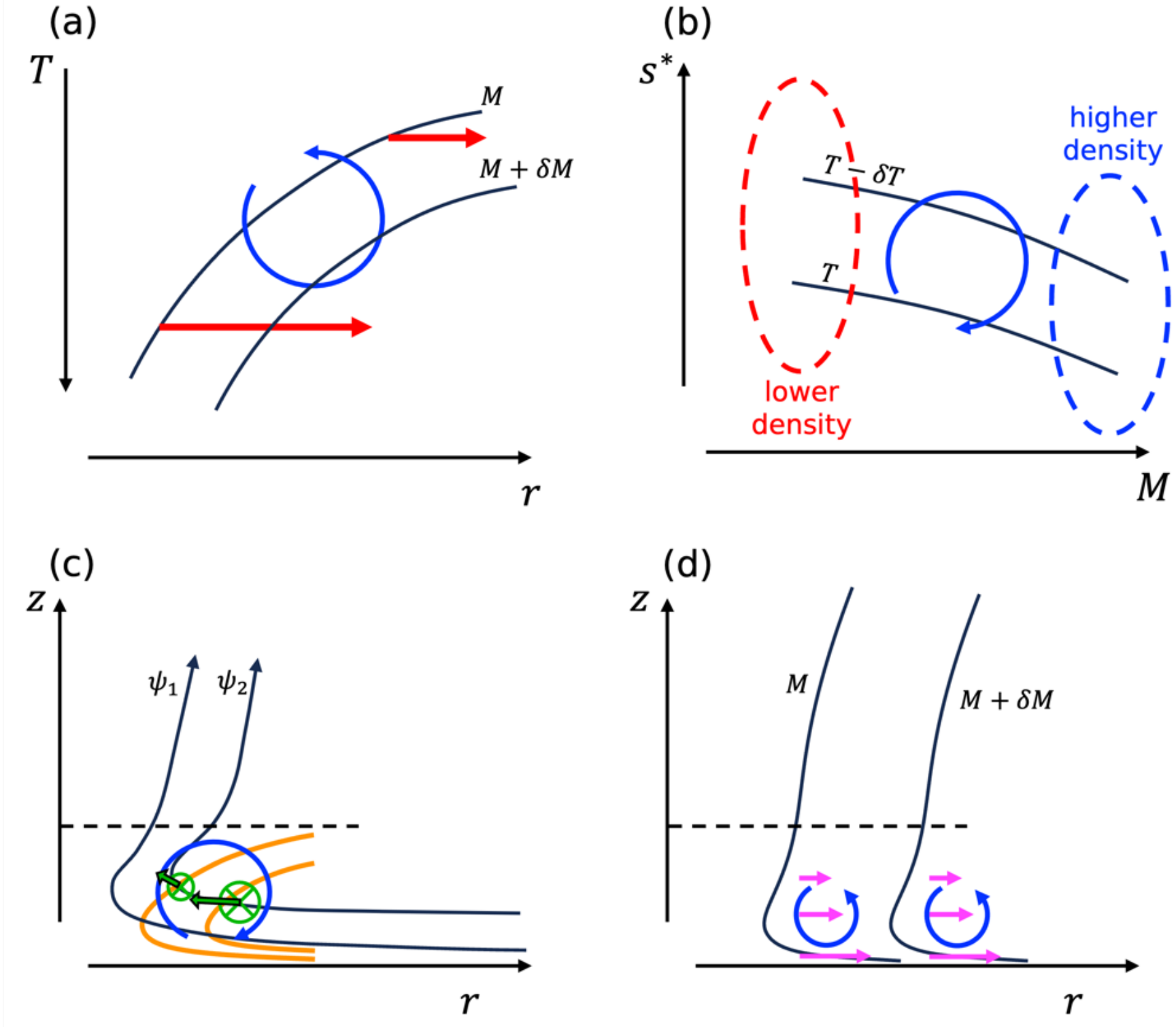

**Fig. 2:** Schematic diagrams illustrating (a) the azimuthal torque (blue circular arrow) of the centrifugal force (red arrows) associated with curved $M$ surfaces (black isolines) in $(r, T)$ space; (b) the baroclinic torque (blue circular arrow) induced by the radial gradient of temperature (and density indicated by red and blue dashed ellipses); (c) the azimuthal torque (blue circular arrow) induced by the convergence of azimuthal vorticity flux (green arrows and cross signs); (d) azimuthal torque (blue circular arrow) induced by radial and vertical friction (magenta arrows). Orange contours in (c) indicate radial velocity isolines. The black dashed line in (c) and (d) indicate the boundary layer top. Both (a) and (b) show centrifugal and baroclinic torque near the upper troposphere where $s^*$ increases vertical upward.

Now we integrate (8) along an $M$ isoline from some temperature $T_b$ at the top of boundary layer ($z_b$) to the potential radius $r_o$ (where $v = 0$ and defines the outflow temperature $T_o$), yielding

$$\int_{T_b}^{T_o} \frac{2M}{r^3} \left(\frac{\partial r}{\partial T}\right)_M dT = \int_{T_b}^{T_o} \left(\frac{\partial s^*}{\partial M}\right)_T dT \,,$$

$$-\frac{M}{r_o^2} + \frac{M}{r_b^2} = \int_{T_b}^{T_o} \left(\frac{\partial s^*}{\partial M}\right)_T dT \,,$$

$$\frac{v_b}{r_b} = \int_{T_b}^{T_o} \left(\frac{\partial s^*}{\partial M}\right)_T dT \,, \tag{9}$$

where $v_b$ denotes the tangential wind at $z_b$. Equation (9) shows that the balanced tangential wind at the low level is proportional to the accumulation of $\left(\frac{\partial s^*}{\partial M}\right)_T$, i.e., $\frac{\partial s^*}{\partial M}$ *evaluated holding* *temperature T constant*, from $T_b$ to $T_o$. Equation (9) is the natural extension of equation (12) of E86 ( $\frac{M}{r^2} = -\frac{ds^*}{dM}(T_b - T_o)$) into the non-SN regime. For a sanity check, we see that under the SN condition, $\left(\frac{\partial s^*}{\partial M}\right)_T = \frac{ds^*}{dM}$ and is only a function of $M$, reducing (9) to $\frac{v_b}{r_b} = -\frac{ds^*}{dM}(T_b - T_o)$.

It should be noted that when symmetric neutrality is not assumed, $\left(\frac{\partial s^*}{\partial M}\right)_T$ is no longer only a function of $M$, but also depends on $T$. The structural evolution of $\left(\frac{\partial s^*}{\partial M}\right)_T$ as a function of $T$ within the eyewall region will be examined in a later section.

In general, both $T_b$ and $T_o$ are functions of $M$. However, for a given $M$ surface, if we first determine the $T_b$ and $T_o$ values of the $M$ surface at the present time and treat them as constants (denoted as $T_b'$ and $T_o'$), we can then further simplify (9) by exchanging the integration and differentiation operators using the Leibniz rule of integration

$$\frac{v_b}{r_b} = \frac{d}{dM}\left(\int_{T_b'}^{T_o'} s^* dT\right) . \qquad (10)$$

Note that at any given time, $s^*$ is a function of $M$ and $T$, so $\int_{T_b'}^{T_o'} s^* dT$ is only a function of $M$ and hence the total derivative. Equation (10) allows the evaluation of $v_b$ without the complexity of computing partial derivatives holding temperature fixed. It also shows that the distribution of $s^*$ along $M$ surfaces directly determines $v_b$ at the low level. Specifically, for a narrow region bounded by two adjacent $M$ surfaces, the difference in the entropy integral $\int_{T_b'}^{T_o'} s^* dT$ between these two $M$ surfaces is proportional to the balanced $v_b$ at the TC BL top. This result has several implications. First, under non-SN conditions, a saturated TC eyewall is *a frontal zone of integrated saturation entropy across M surfaces*, a generalization of the findings derived by Emanuel (1997). Second, equation (10) indicates that any process that reduces the gradient of the integrated saturation entropy across M surfaces will weaken the balanced intensity. One such process is ventilation, where environmental dry air is injected into the TC eyewall updraft. Despite assuming saturation, the above inference does not rely on the SN assumption, which was one of the core assumptions of the ventilation theory of Tang and Emanuel (2009).

The derivations of (9) and (10) use the equivalent forms of a Jacobian matrix determinant (Appendix A). As shown in Appendix B, it is possible to derive (9) and (10) entirely based on the energetic (integral) property (5) alone.

*b. General form*

A more general derivation of the tangential wind formula in height coordinate, which includes the nonlinear advection of radial and vertical momentum and boundary layer friction, is provided in Appendix C. The inclusion of these processes results in additional terms:

$$\frac{v_b}{r_b} = \int_{T_b}^{T_o} \left(\frac{\partial s^*}{\partial M}\right)_T dT + \int_{\tilde{\eta}_b/r_b}^{\tilde{\eta}_o/r_o} \left(\frac{\partial \psi}{\partial M}\right)_{\tilde{\eta}/r} d\left(\frac{\tilde{\eta}}{r}\right) - \int_{r_b}^{r_o} \left(\frac{\partial F_r}{\partial M}\right)_r dr - \int_{z_b}^{z_o} \left(\frac{\partial F_z}{\partial M}\right)_z dz\,, \qquad (11)$$

where $F_r$ and $F_z$ are the radial and vertical momentum forcings, $\psi$ is the nondivergent streamfunction, and $\tilde{\eta} = \frac{1}{\rho}\left(\frac{\partial u}{\partial z} - \frac{\partial w}{\partial r}\right)$ with $\rho$ the density of moist air. It should be noted that since gradient wind balanced is no longer assumed, the path integration in (11) extends into the TC BL, where saturation assumption is satisfied. All variables in (11) with subscript "$b$" denote evaluation at level $z_b$ within the TC BL. Equation (11) summarizes the different processes contributing to the azimuthal torque balance to maintain the curved structure of $M$ surfaces and the low-level wind intensity of a saturated non-SN vortex. As explained in appendix C and illustrated in Fig. 2, in addition to the balanced term, the second term of the R.H.S represents the effect of azimuthal torque induced by azimuthal vorticity flux convergence (Figs. 2c), whereas the sum of third and fourth terms represent the effect of frictional torque (Figs. 2d).

If we now focus on the 2nd term and assume steady state and symmetric neutrality above the boundary layer (where $F_r = F_z = 0$), $\left(\frac{\partial \psi}{\partial M}\right)_{\tilde{\eta}/r}$ then becomes $\frac{d\psi}{dM}$ and is only a function of $M$. Therefore, $\int_{\tilde{\eta}_b/r_b}^{\tilde{\eta}_o/r_o} \left(\frac{\partial \psi}{\partial M}\right)_{\tilde{\eta}/r} d\left(\frac{\tilde{\eta}}{r}\right) = \frac{d\psi}{dM}\left(\frac{\tilde{\eta}_o}{r_o} - \frac{\tilde{\eta}_b}{r_b}\right) \approx -\frac{d\psi}{dM}\frac{\tilde{\eta}_b}{r_b}$, which is precisely the unbalanced term derived by BR09b (their equation 19).

BR09b neglected boundary layer friction in deriving their equation (20). This is because, at steady state, SN condition is true only above the boundary layer where $F_r$ and $F_z$ vanish. Since the SN assumption is now relaxed in the present study, the derivation of (11) reveals the contributions of boundary layer frictions and can be applied within the TC boundary layer. It should, therefore, be noted that it is the sum of the $\tilde{\eta}$ term (2nd term) and boundary layer friction term (3rd and 4th terms) in (11) together that represents the unbalanced, supergradient wind component (Montgomery and Smith 2017; Rotunno 2022).

*c. A correction term to the balanced component*

As discussed in BR09a, E86 neglected the contribution of vapor density to the balanced intensity. The error induced by this approximation on the balanced intensity can be corrected by the following correction term, as derived in appendix D

$$\left(\frac{v_b}{r_b}\right)_{correction} = -\int_{T_b}^{T_o} \gamma \left(\frac{\partial s^*}{\partial M}\right)_T dT \, , \tag{12}$$

where $\gamma = \left(\frac{1+\frac{L_v}{R_v T}}{1+\frac{L_v q^*}{R_d T}}\right) q^*$ and $R_v$ is the gas constant for water vapor. The above correction shares some similarities to Emanuel (1988) but is different in that the total water content is not assumed to be a function of $M$ alone. Combining (12) with (11), the final form of the tangential wind formula is

$$\frac{v_b}{r_b} = \int_{T_b}^{T_o} \left(\frac{\partial s^*}{\partial M}\right)_T (1-\gamma) dT + \int_{\tilde{\eta}_b/r_b}^{\tilde{\eta}_o/r_o} \left(\frac{\partial \psi}{\partial M}\right)_{\tilde{\eta}/r} d\left(\frac{\tilde{\eta}}{r}\right) - \int_{r_b}^{r_o} \left(\frac{\partial F_r}{\partial M}\right)_r dr - \int_{z_b}^{z_o} \left(\frac{\partial F_z}{\partial M}\right)_z dz \, , \tag{13}$$

where the first term is referred to as the bias-corrected balanced wind component. If the SN condition is assumed, $\left(\frac{\partial s^*}{\partial M}\right)_T$ becomes $\frac{ds^*}{dM}$ and is constant along the integration path. Further assuming $M \approx r_b v_b$, the bias-corrected maximum potential intensity (MPI) under SN condition is

$$\begin{aligned} {v_b}_{MPI}^2 &= M \frac{ds^*}{dM} \int_{T_b}^{T_o} (1-\gamma) dT \, , \\ &= -(\overline{1-\gamma}) M \frac{ds^*}{dM} (T_b - T_o) \, , \end{aligned} \tag{14}$$

where $\overline{1-\gamma} \equiv \frac{1}{T_o - T_b} \int_{T_b}^{T_o} (1-\gamma) dT$. If we define $v_{E86}^2 = -M \frac{ds^*}{dM}(T_b - T_o)$, the bias-corrected maximum intensity is ${v_b}_{MPI}^2 = (\overline{1-\gamma}) v_{E86}^2$.

## 3. Methods of evaluation

*a. Numerical model simulations*

To verify the results in section 2, we use the Cloud Model 1 (CM1; Bryan and Fritsch 2002), version 20.1, to conduct numerical simulations using the “default” axisymmetric TC setting. This setting is similar to the second configuration in Bryan (2012), except that a modified Rankine vortex profile is used:

$$v_0(r) = \begin{cases} v_m\left(\frac{r}{r_{m1}}\right) & r < r_{m1} \\ v_m\left(\frac{r}{r_{m1}}\right)^{-0.35} & r_{m1} \leq r < r_{m2} \\ w_2 \times v_m\left(\frac{r}{r_{m1}}\right)^{-0.35} + w_3 \times \frac{v_m}{2}\left(1 - \frac{r - r_{m2}}{r_{m3} - r_{m2}}\right) & r_{m2} \leq r < r_{m3} \\ 0 & r \geq r_{m3} \end{cases}, \quad (15)$$

where $v_m = 15\ ms^{-1}$ and $r_{m1} = 75\ km$ are the initial maximum tangential wind and radius of maximum wind; $r_{m2} = 200\ km$ and $r_{m3} = 500\ km$; $w_3 := \frac{r - r_{m2}}{r_{m3} - r_{m2}}$ and $w_2 := 1 - w_3$. Given the surface wind profile, the full tangential wind is obtained with $v_m$ decaying linearly from 15 $ms^{-1}$ at the surface to 0 $ms^{-1}$ at z = 15 km. The moist tropical sounding (Dunion 2011) and a sea surface temperature of 28 °C are used to initialize the simulations. To better conserve saturation entropy, the Rotunno-Emanuel (1987) simple water-only scheme is used.

The total domain size of the simulation is 1500 km × 25 km. Horizontal grid spacing $\Delta r = 500\ m$ is used within a 350 km radius, beyond which $\Delta r$ is stretched to a maximum value of 4 km at r = 1500 km. A variable vertical grid spacing is used below $z = 5375\ m$ with a maximum $\Delta z = 200\ m$, yielding a total of 141 vertical levels.

Following the suggested setting of Bryan (2012), $C_E$ is set to $1.2 \times 10^{-3}$ (Drennan et al. 2007), while we use the "default" CM1 setting for $C_D$, defined based on Fairall et al. (2003) at low wind conditions and capped at $2.4 \times 10^{-3}$. Similar to previous studies, we do not use a realistic radiation scheme but include the Newtonian cooling term in the potential temperature equation that mimics the radiative cooling effect, which is capped at 2 K $\text{day}^{-1}$.

Because of the large variability during the TC intensification, a set of 11-member ensemble simulations are performed to remove the random fluctuations in the intensification process. First, a control simulation (CTRL) is performed using the setting described above. At hour 21, random perturbations of the potential temperature of 0.1 K magnitude are added to the inner-core region with $r \leq 100$ km and $z \leq 15$ km to create 10 perturbed simulations, forming the 11-member ensemble. Perturbations are added at hour 21 instead of at the initial condition because this time is nearly at the end of the boundary layer spin-up process and is before the onset of the intensification process. As such, all perturbed members intensify with similar onset timing of RI and intensification rate as the CTRL.

*b. Evaluation of path integrals*

One additional challenge is that $\left(\frac{\partial s^*}{\partial M}\right)_T$ (or $\left(\frac{\partial \psi}{\partial M}\right)_{\tilde{\eta}/r}$) may have singularities along the integration path. Fortunately, we showed in Appendix E that these singularities can always be regularized such that the integrals in (11) are always integrable. Specifically, the following expressions are proven to be the optimal regularizations for $\left(\frac{\partial s^*}{\partial M}\right)_T$ and $\left(\frac{\partial \psi}{\partial M}\right)_{\tilde{\eta}/r}$

$$\widetilde{\left(\frac{\partial s^*}{\partial M}\right)_T} \equiv \frac{1}{\delta T}\int_{l_i-\frac{\delta l}{2}}^{l_i+\frac{\delta l}{2}} J\left(\frac{s^*, T}{r, z}\right)\frac{dl}{|\nabla M|}\,, \tag{16a}$$

$$\widetilde{\left(\frac{\partial \psi}{\partial M}\right)_{\tilde{\eta}/r}} \equiv \frac{1}{\delta(\tilde{\eta}/r)}\int_{l_i-\frac{\delta l}{2}}^{l_i+\frac{\delta l}{2}} J\left(\frac{\psi, \tilde{\eta}/r}{r, z}\right)\frac{dl}{|\nabla M|}\,, \tag{16b}$$

where $l_i$ is the i-th grid point along the desired $M$ surface (parameterized by path length $l$). When singularity does not occur, $\widetilde{\left(\frac{\partial s^*}{\partial M}\right)_T}$ and $\widetilde{\left(\frac{\partial \psi}{\partial M}\right)_{\tilde{\eta}/r}}$ equal their averaged values between $[l_i - \frac{\delta l}{2}, l_i + \frac{\delta l}{2}]$. When singularity occurs within $[l_i - \frac{\delta l}{2}, l_i + \frac{\delta l}{2}]$, $\widetilde{\left(\frac{\partial s^*}{\partial M}\right)_T}$ and $\widetilde{\left(\frac{\partial \psi}{\partial M}\right)_{\tilde{\eta}/r}}$ enable the accurate integration across the singularity. Because (16a,b) can be used regardless of the presence of singularity, they will be used to compute the first and second terms of (13) in the subsequent analysis.

## 4. Results

### *a. Timing of peak intensification rate and intensification period*

Since the main focus of the analysis is on the intensification period, we first identify the timing of the maximum intensification rate ($t_{RI}$). For each ensemble member, a 12-hourly running average is applied to the maximum tangential wind ($v_{max}$) time series (Fig. 3a) to remove the short-term intensity fluctuations. The time derivative of the smoothed $v_{max}$ time series is then computed to determine the $t_{RI}$ of each ensemble member. The intensification period of each member is defined as the 24-hour period centered at $t_{RI}$. As shown in Figs. 3b,c, when composited relative to $t_{RI}$, the ensemble-mean intensification rate shows a clean peak at $t_{RI}$ with persistent increase in $v_{max}$ over the 24-hour window.

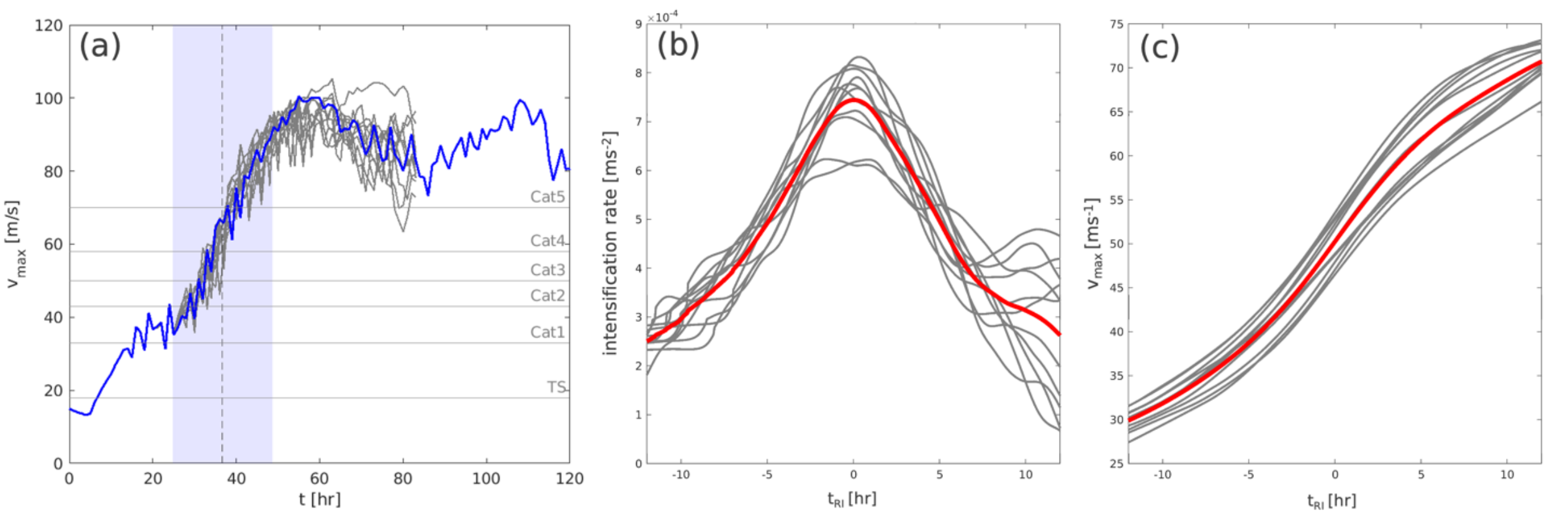

**Fig. 3:** (a) Time series of the $v_{max}$ of the CTRL (blue) and perturbed (grey) members. The 24-hour time window centered at the $t_{RI}$ of the CTRL member is highlighted in blue. (b) Composite of 12-hourly smoothed intensification rate of all members with respect to their $t_{RI}$. The composite mean is shown in red, while each ensemble member is shown in grey. (c) is the same as (b), but for 12-hourly smoothed $v_{max}$ time series.

*b. Evaluation of the generalized tangential wind formula*

We now focus on evaluating (13) during the 24-hour intensification time period. For a given time $t$ during the intensification period, a 3-hour time averaging centered at $t$ will be applied to all physical fields to remove high-frequency fluctuation. These temporally-averaged fields are then used to determine the appropriate $M$ surfaces and the computation of each term in equation (13). It is important to note that the $J\left(\frac{\psi,\tilde{\eta}/r}{r,z}\right)$ term in (16b) is nonlinear, and therefore is first computed using instantaneous fields and then temporally averaged.

Given that the derivation of the balanced wind component (9) relies on saturation condition and that diabatic heating release is an essential forcing that drives TC secondary circulation and intensification, the set of $M$ surfaces are chosen to maximize the amount of diabatic heating along the $M$ surfaces. Firstly, the averaged density-weighted diabatic heating between z = 2 km and the outflow radius is computed for all $M$ surfaces at the vicinity of the the eyewall updraft

$$Q_M = \frac{\int_{z=2\,km}^{z=z_o} \rho Q\, dl}{\int_{z=2\,km}^{z=z_o} dl}\ , \tag{17}$$

where $Q$ is the diabatic forcing from microphysics scheme. The $M$ surface with the maximum $Q_M$ is chosen ($M_{maxQ}$). Then, 50 $M$ surfaces within the $\pm$ 1 km radial range of the $M_{maxQ}$ surface at z = 2 km are sampled. To ensure the validity of the saturation condition, we compute the averaged relative humidity deficit ($RH_{deficit}$) along the 50 $M$ surfaces

$$RH_{deficit} = \frac{\int_{z=750m}^{z=z_o} (100\% - RH)\, dl}{\int_{z=750m}^{z=z_o} dl}\ , \tag{18}$$

where $RH$ is the relative humidity. $M$ surfaces that have $RH_{deficit} \geq 2\%$ are discarded.

Fig. 4 shows the distribution of the selected $M$ surfaces (gray) at the time of $t_{RI}$. It is noted that the majority of the selected $M$ surfaces that pass through the saturated diabatic heating region are located at the slight inward side of the maximum tangential wind jet core (Fig. 4b), consistent with the knowledge that TC intensification is most efficient when diabatic heating occurs inside of the radius of maximum wind (Schubert and Hack 1983). In contrast, the $M$ surface that passes through the maximum tangential wind jet core (blue) becomes strongly subsaturated immediately above 2 km and is located at the outer edge of the diabatic heating region, with an averaged $Q_M = 4.6 \times 10^{-4}\ \mathrm{kg\ K\ m^{-3}\ s^{-1}}$, which is about 60% of the selected $M$ surfaces ($Q_M = 7.6 \times 10^{-4}\ \mathrm{kgKm^{-3}s^{-1}}$). For this reason, the analyses in subsequent sections focus on the selected $M$ surfaces that maximize $Q_M$ and minimize $RH_{deficit}$.

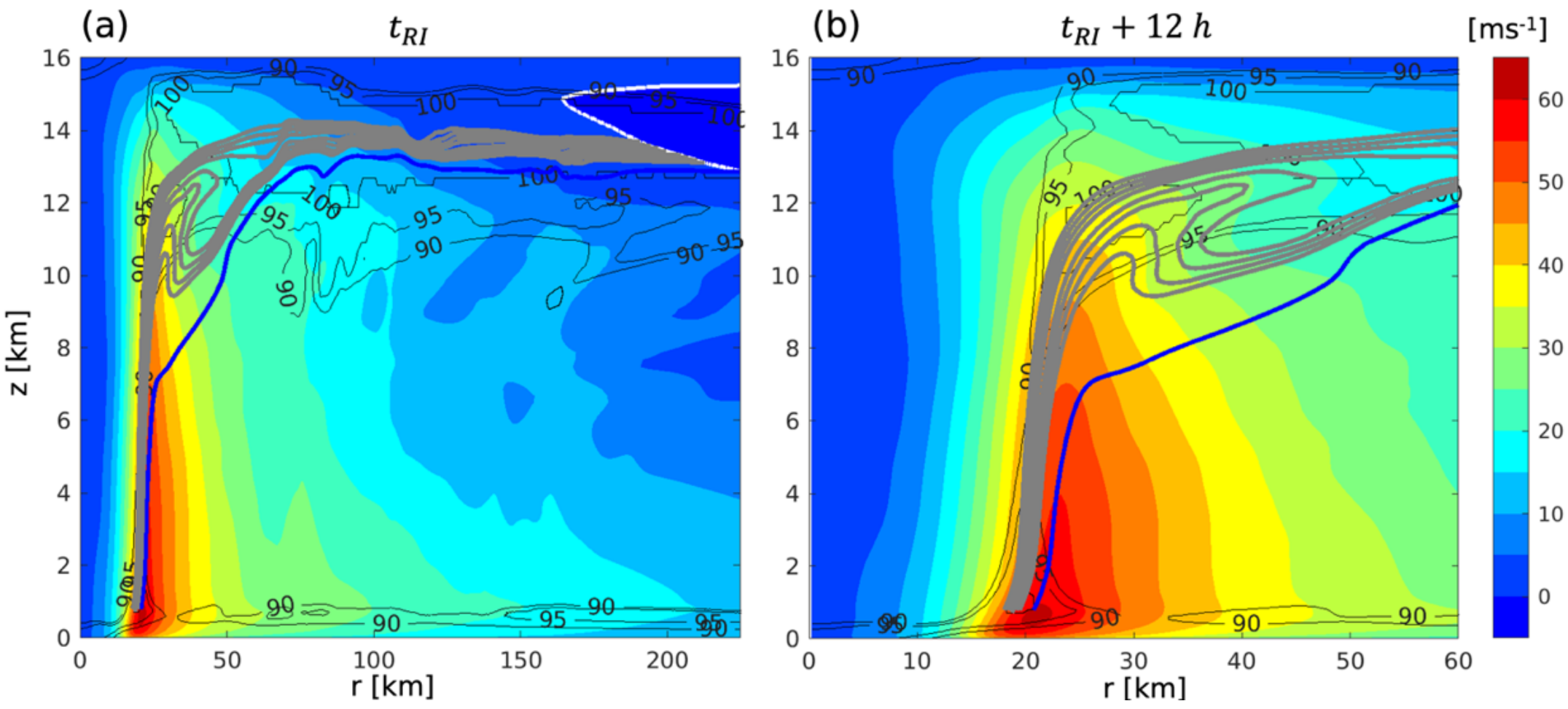

**Fig. 4:** Families of $M$ surfaces of the CTRL experiment that satisfy the selection criteria (see text for details) at $t_{RI}$. Thin grey contours are relative humidity and white line denotes contour where $v = 0$ m s-1. (b) is similar to (a) but zooms in to display the details of the selected $M$ surfaces at the eyewall region.

We now evaluate equation (13) within the intensification period, as shown in Fig. 5. The result shows that the sum of all terms (red) almost perfectly agrees with the actual $v_b$ throughout the entire intensification period. Importantly, previous studies showed that applying the traditional formula from PI (i.e., $\frac{v_b}{r_b} = -\frac{ds^*}{dM}(T_b - T_o)$) to the intensification period can result in rather noisy data, particularly during the early stage of the intensification period (see Fig. 1 of Peng et al. 2018).

Our current result from Fig. 5 not only confirms that the accuracy of equation (13), but it also shows that the generalized terms in equation (13) are the proper and natural extension of previous PI formulations (Emanuel 1987 and Bryan and Rotunno 2009) into the non-SN vortex regime, which is applicable to the entirety of the intensification period.

Among the terms, the bias-corrected balanced component is the dominant term that shows the largest increase during the intensification period. It is interesting to note that the bias term (magenta dashed line) is close to zero before $t_{RI}$ but decreases to about -5 m s$^{-1}$ near the end of the intensification period, which can result in noticeable positive bias in the diagnosed $v_b$ if the correction term is not included.

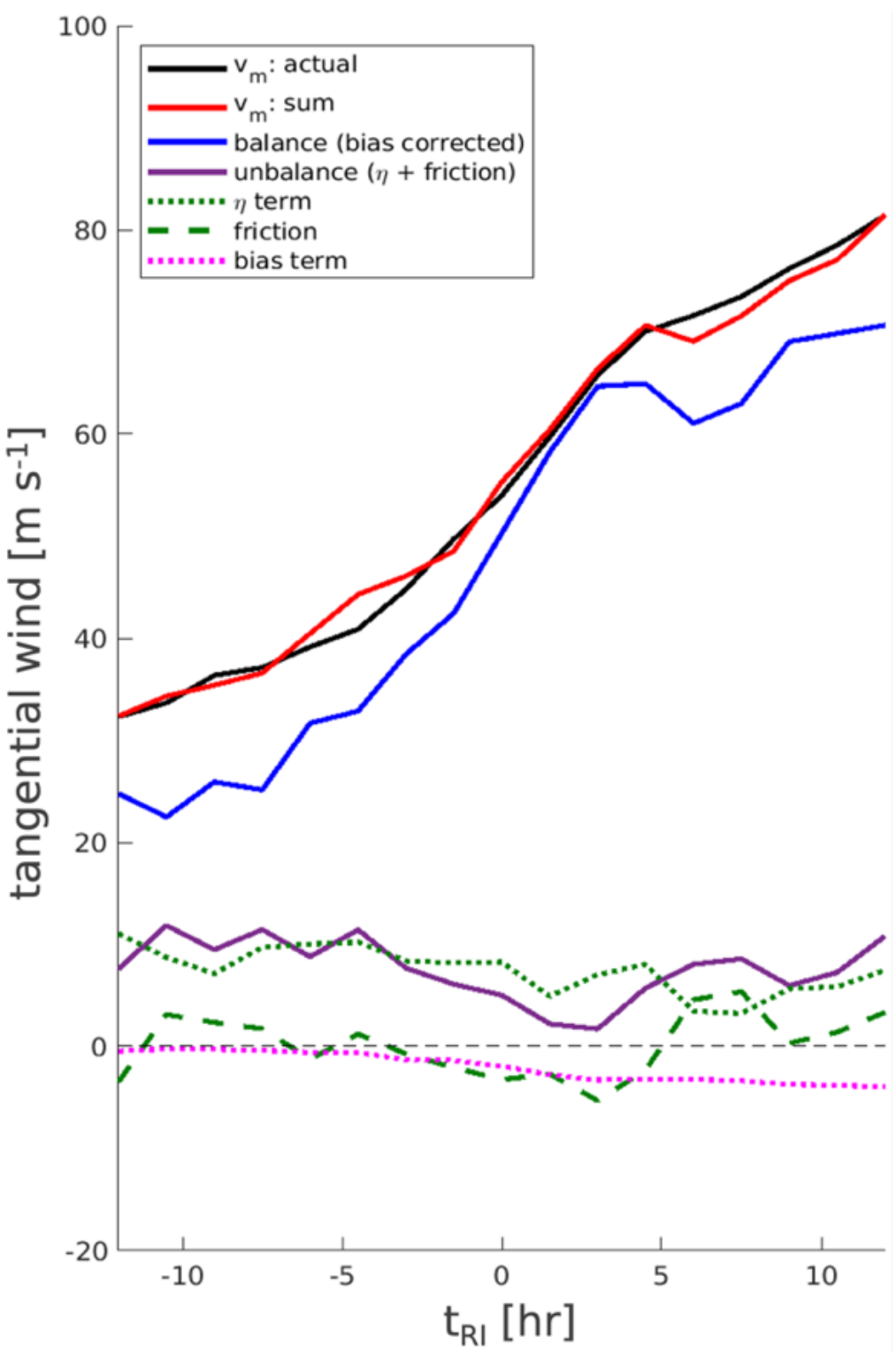

**Fig. 5:** Composite-mean time series for each term in equation 13. The actual $v_m$ is shown in black. The solid purple line shows the unbalanced contribution, which is the sum of the $\eta$ term (green dotted line) and frictional contributon (green dashed line).

On the other hand, the unbalanced term (sum of $\eta$ term and frictional term) remains nearly constant during the intensification period. Peng et al. (2018) found that the evolution of the unbalanced term appears to supplement the difference in the intensification rate between the actual maximum tangential wind and $v_{EPI}$, which contrasts with our current results. One possible reason is that our experiment uses a horizontal mixing length of 1500 m (default setting from CM1), while Peng et al. 2018 used a rather small mixing length of 94 m, which can dramatically increase the supergradient component. With our current model setting, the result in Fig. 5 shows that the significant increase in $v_b$ is largely attributed to the spin-up of the balanced vortex. Therefore, in the subsequent analyses, we only focus on examining the balanced term. We note here that even though the unbalanced term is small and largely uniform, it could potentially play important role in the intensification process (Smith et al. 2008), which deserves further investigation in future studies.

*c. Dependence of $\left(\frac{\partial s^*}{\partial M}\right)_T$ on temperature*

Equation (13) shows that for a non-SN vortex, the dependence of $\left(\frac{\partial s^*}{\partial M}\right)_T$ on $T$ is important to the balanced tangential wind at $z_b$. In this section, we examine this dependence closely.

Several interesting structures of $\left(\frac{\partial s^*}{\partial M}\right)_T$ emerge when we examine the composite-mean evolution of $\left(\frac{\partial s^*}{\partial M}\right)_T$, as shown in Fig. 6. First, at $t_{RI}$, $\left(\frac{\partial s^*}{\partial M}\right)_T$ shows a clear linear structure as a function of $T$, with large negative value near 210-240 K and decaying towards the low level (Fig. 6e). Consistent with previous studies (e.g., Peng et al. 2018, 2019), this linear structure of $\left(\frac{\partial s^*}{\partial M}\right)_T$ confirms that the SN assumption is not valid during the early intensification period (including $t_{RI}$), since $\left(\frac{\partial s^*}{\partial M}\right)_T$ would otherwise be *constant* with respect to variation of $T$ along the $M$ surfaces.

At $t_{RI} + 3h$, the large negative value of $\left(\frac{\partial s^*}{\partial M}\right)_T$ descend towards lower levels near 240 K (Fig. 6f). This trend continues in the subsequent hours, and $\left(\frac{\partial s^*}{\partial M}\right)_T$ becomes more uniform in the mid-troposphere near 220 K - 260 K (Fig. 6g-i), indicating the vortex in the free troposphere becomes more symmetric neutral at this late stage, which agrees with E12 and Peng et al. (2018), (2019).

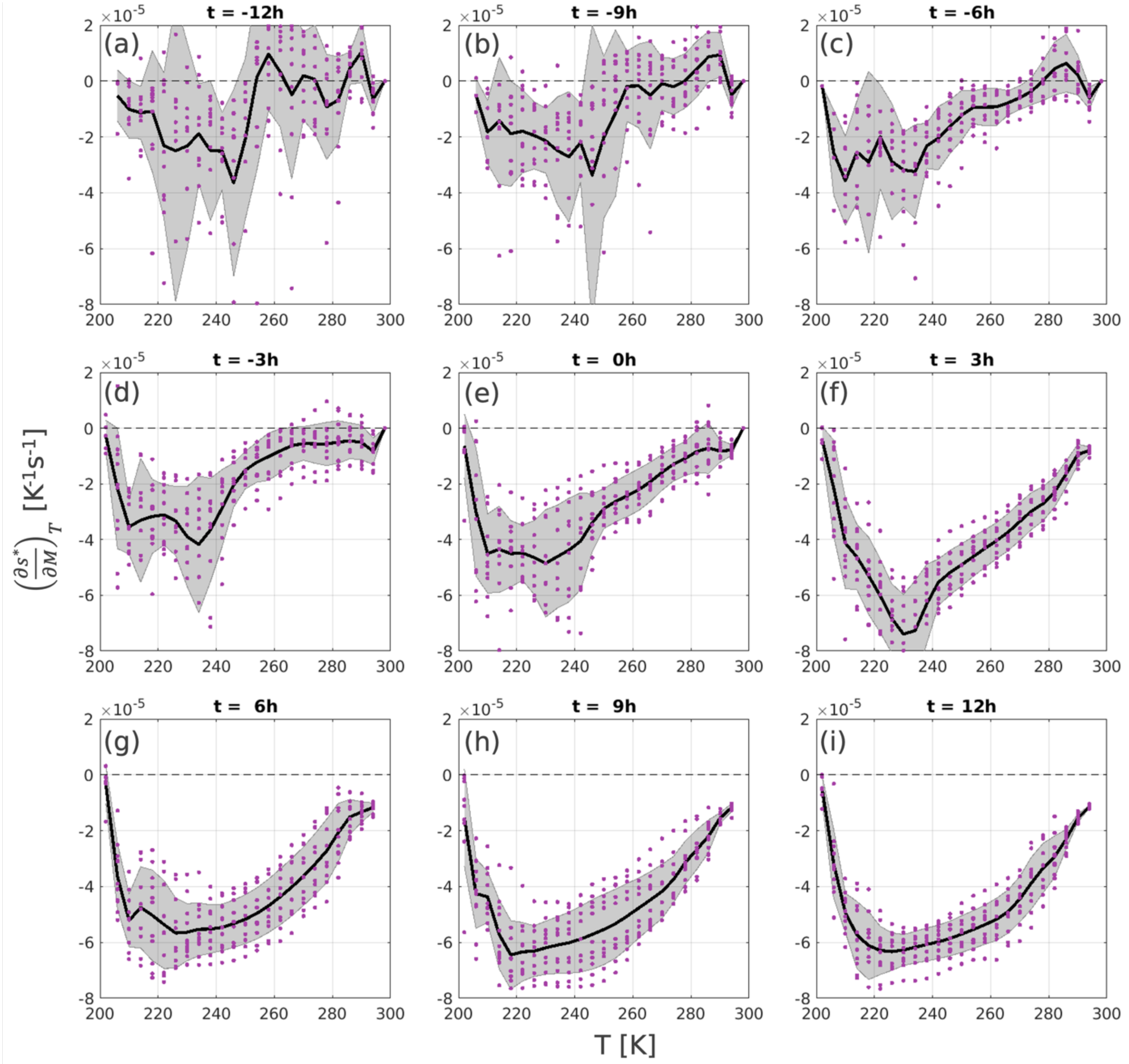

**Fig. 6:** Profiles of $\left(\frac{\partial s^*}{\partial M}\right)_T$ as a function of $T$ at every three hours during the intensification time window. The Black lines are the composite mean profile with shaded region indicate $\pm 1$ standard deviation. The data points of each member are shown by the purple dots.

Another important and interesting feature emerges when compositing the mean $\left(\frac{\partial s^*}{\partial M}\right)_T$ profiles during the intensification period, as shown in Fig. 7a. Near $t_{RI} - 6\ h$, $\left(\frac{\partial s^*}{\partial M}\right)_T$ begins to develop a large negative value near 210-240 K, which later further strengthens to $-4 \times 10^{-5}\ K^{-1} s^{-1}$ at $t_{RI} - 3h$ and $t_{RI}$. As shown in equation (8), $\left(\frac{\partial s^*}{\partial M}\right)_T$ is proportional to the local slope of the $M$ surface (e.g., $\left(\frac{\partial r}{\partial T}\right)_M$), with large value corresponding to a strong inward bending of the $M$ surface. Therefore, a top-heavy profile of $\left(\frac{\partial s^*}{\partial M}\right)_T$ implies that the majority of the inward bending of the $M$ surfaces is concentrated at the upper troposphere, indicating the development of a deep tall vortex that reaches the tropopause during $t_{RI} - 6h$ to $t_{RI}$. This result suggests that the development a deep tall vortex preconditions the onset of maximum intensification rate, which is supported by the recent observational evidence on the relationship between vortex depth and intensification onset (Fischer et al. 2022, 2025; DesRosiers, et al. 2023).

At steady state after t = 8 days (Fig. 7b), $\left(\frac{\partial s^*}{\partial M}\right)_T$ indeed becomes nearly uniform at $-6 \times 10^{-5}\ K^{-1}s^{-1}$ throughout the troposphere, while the boundary layer ($T > 280\ K$) shows noticeable gradient. This result confirms that symmetric neutrality is indeed a good approximation for steady-state TCs above the boundary layer. It is interesting to see that at steady state, the $M$ surfaces reach the potential radius at a temperature of about 208 K, which is higher than the outflow temperature during the intensification period. This could potentially be due to the self-stratification process examined in ER11.

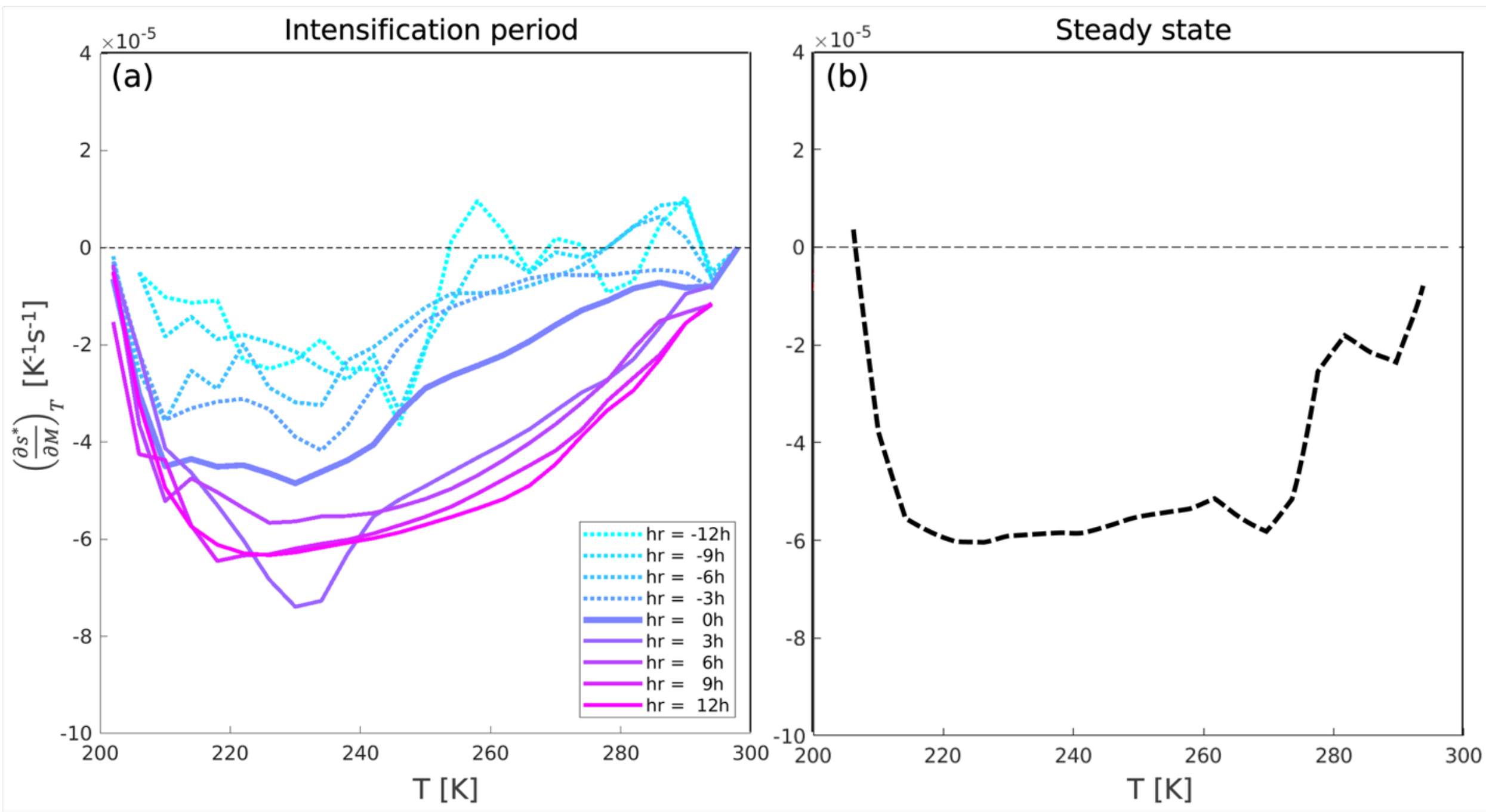

**Fig.7:** (a) Evolution of the composite mean profiles of $\left(\frac{\partial s^*}{\partial M}\right)_T$ as a function of $T$ at every three hours during the intensification time window. The profiles at $t_{RI}$ (hr = 0 h) is thickened. (b) Similar to (a), but for steady state period during t = 192-198 hour of the CTRL member.

Why does $\left(\frac{\partial s^*}{\partial M}\right)_T$ develop a large negative value near the tropopause, and what processes control its value there? To answer these questions, we write $\left(\frac{\partial s^*}{\partial M}\right)_T$ as $\left(\frac{\partial s^*}{\partial M}\right)_T = \left(\frac{\partial s^*}{\partial M}\right)_p + \left(\frac{\partial s^*}{\partial p}\right)_M \left(\frac{\partial p}{\partial M}\right)_T$. Because the second term is much smaller in the outflow region, we have $\left(\frac{\partial s^*}{\partial M}\right)_T \approx \left(\frac{\partial s^*}{\partial M}\right)_p = \left(\frac{\partial s^*}{\partial r}\right)_p / \left(\frac{\partial M}{\partial r}\right)_p$, which is proportional to baroclinicity divided by inertial stability. The cross-section of $\left(\frac{\partial s^*}{\partial M}\right)_p$ of the unperturbed member at $t_{RI}$ (Fig. 8a) shows that large negative values of $\left(\frac{\partial s^*}{\partial M}\right)_p$ exist at temperature from 205 K to 210 K and radii from 30 to 60 km, which is due to a combination of negative $\left(\frac{\partial s^*}{\partial r}\right)_p$ (Fig. 8b) and large positive $\left(\frac{\partial M}{\partial r}\right)_p^{-1}$ at these radii (Fig. 8c). Therefore, the decrease of inertial stability at the outflow region is the major cause of the large negative values of $\left(\frac{\partial s^*}{\partial M}\right)_T$ near 205-210 K.

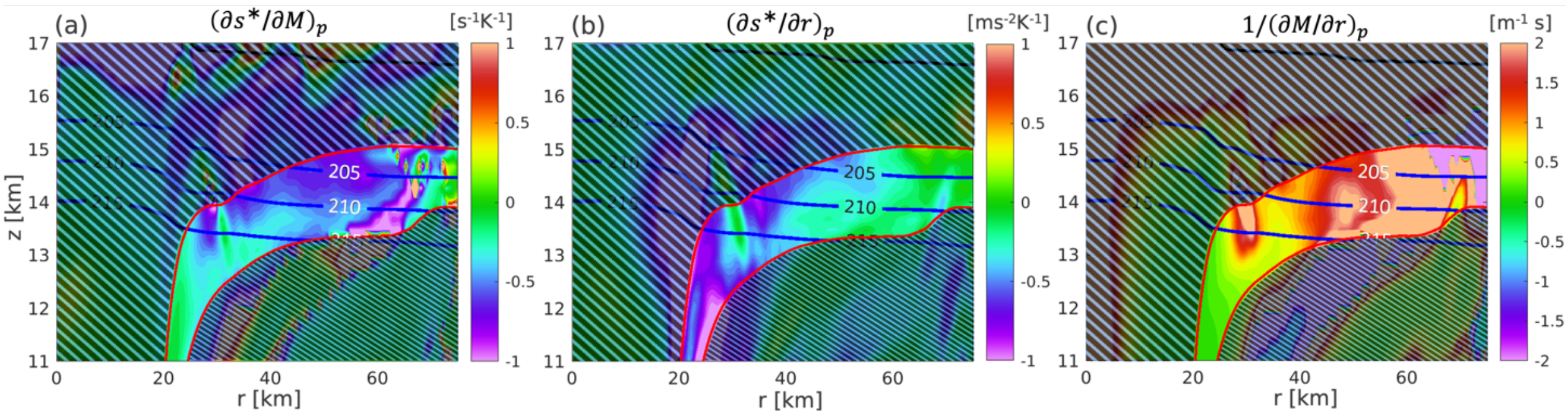

**Fig.8:** (a) Cross-sections of $\left(\frac{\partial s^*}{\partial M}\right)_p$ normalized by $10^{-4}$ of the unperturbed member at $t = t_{RI}$. The regions outside the smallest and largest $M$ surfaces (red contours) are hatched. Blue contours indicate temperature contours of 205, 210 and 215 K. (b) and (c) are the same as (a), but for cross-sections of $\left(\frac{\partial s^*}{\partial r}\right)_p$ and $\left(\frac{\partial M}{\partial r}\right)_p^{-1}$, respectively.

The decrease of $\left(\frac{\partial M}{\partial r}\right)_p$ to small value near the outflow region can be understood using the conservation principle of moist potential vorticity (PV), $P_e \equiv \frac{1}{r}\frac{(\nabla\theta_e \times \nabla M)\cdot\hat{\lambda}}{\rho}$. In the free troposphere of an axisymmetric flow where the forcing of angular momentum and moist entropy are small (assuming reversible moist process), $P_e$ is approximately conserved due to the vanishing baroclinic term under axisymmetric or saturation condition[3] (both satisfied in the present case). Under non-SN conditions, $P_e$ at the low-level eyewall region is dominantly negative, owing to the positive vertical vorticity and decrease of $\theta_e$ with height (i.e., convective instability). As the parcels rise along the slantwise eyewall updraft to the upper tropospheric outflow region, the vertical gradient of $\theta_e$ changes sign from negative to positive as $\theta_e$ asymptotes to $\theta$ at the upper troposphere. To conserve the negative moist-PV, there must be a mechanism to change the parcel's vertical vorticity (and dry PV, $P \equiv \frac{1}{r}\frac{(\nabla\theta\times\nabla M)\cdot\hat{\lambda}}{\rho}$) from positive to negative along the ascending path.

The mechanism driving the required sign change of vorticity (and dry PV) can be shown more explicitly by rewriting the moist-PV conservation principle in terms of the dry PV, i.e., $\frac{DP}{Dt} = \frac{\zeta_a\cdot\nabla Q}{\rho}$ where $Q \approx -\frac{L_v\dot{q}}{\theta C_p T}$. Since $Q$ is spatially correlated with vertical motion $w$, the forcing term is proportional to $\zeta_z\frac{\partial w}{\partial z} + \boldsymbol{\zeta}_h\cdot\nabla w$, both of which are negative at the outflow level (e.g., vorticity compression and negative tilting). Therefore, the moist-PV conservation dictates that the diabatic heating of the convectively unstable parcels throughout the eyewall must ultimately result in the emergence of a widespread region of small $\left(\frac{\partial M}{\partial r}\right)_p$ values at the upper-level outflow region.

While the conservation principle of moist PV explains the emergence and development of large negative values of $\left(\frac{\partial s^*}{\partial M}\right)_T$ near the tropopause, it is not sufficient to determine the precise value there. To answer this question, we can use the angular momentum and entropy equations. In pressure coordinates, we write the secondary circulation in terms of nondivergent streamfunction, i.e., $u = -\frac{1}{r}\frac{\partial\psi}{\partial p}$ and $\omega = \frac{1}{r}\frac{\partial\psi}{\partial r}$. Taking time averaging of 3 hours and separating the nonlinear advection terms into mean and transient eddy contributions, the conservation principles of $M$ and $s^*$ can be written as

[3] Schubert et al. (2001) showed that the baroclinic term $\frac{1}{\rho^3}\nabla\theta_e\cdot(\nabla\rho\times\nabla p)$ does not vanish if $\theta_e$ is used to define moist PV. However, the baroclinic term indeed vanishes if the flow is either saturated or axisymmetric. For saturated flow, $\theta_e$ becomes $\theta_{es}$ and is only a function of $\rho$ and $p$, which causes $\theta_e\cdot(\nabla\rho\times\nabla p)$ to vanish. For axisymmetric flow, $\nabla\rho\times\nabla p$ points at the azimuthal direction, while $\nabla\theta_e$ lies in the radius-height plane, so $\nabla\theta_e\cdot(\nabla\rho\times\nabla p)$ also vanishes. Since both of these conditions are satisified in axisymmetric hurricane eyewall, moist PV is close to conserved within eyewall convection in the present study.

$$\frac{\delta M}{\delta t}+\bar{u}\left(\frac{\partial \bar{M}}{\partial r}\right)_{\psi}+\overline{u'\left(\frac{\partial M}{\partial r}\right)_{\psi}'}=\overline{F_M} \ , \tag{19a}$$

$$\frac{\delta s^*}{\delta t}+\bar{u}\left(\frac{\partial \overline{s^*}}{\partial r}\right)_{\psi}+\overline{u'\left(\frac{\partial s^*}{\partial r}\right)_{\psi}'}=\overline{F_{s^*}} \ , \tag{19b}$$

where overbar denotes 3-hourly time means; the ( $'$ ) terms denote temporal perturbations from the 3-hourly time mean; and $\delta$ denotes the change in the 3-hour time interval. The first tendency terms in (19a,b) are at least one order smaller than the other terms and can be neglected. Rearranging the terms and dividing the two equations then yields

$$\left(\frac{\partial \overline{s^*}}{\partial \bar{M}}\right)_{\psi}\approx\frac{\overline{F_{s^*}}+F_{s^*,eddy}}{\overline{F_M}+F_{M,eddy}}; \tag{20}$$

$$F_{s^*,eddy}=-\overline{u'\left(\frac{\partial s^*}{\partial r}\right)_{\psi}'}\ ;\ F_{M,eddy}=-\overline{u'\left(\frac{\partial M}{\partial r}\right)_{\psi}'}\ .$$

The eddy forcing terms ($F_{s^*,eddy}$ and $F_{M,eddy}$) represent the time-mean effects of the resolved nonlinear advection of $s^*$ and $M$ by the transient eddies. As the secondary circulation reaches the tropopause, a large portion of the isolines of the streamfunction spread outward and become nearly parallel to temperature contours, and thus $\left(\frac{\partial \overline{s^*}}{\partial \bar{M}}\right)_{\psi}\approx\left(\frac{\partial \overline{s^*}}{\partial \bar{M}}\right)_{T}$ and

$$\left(\frac{\partial \overline{s^*}}{\partial \bar{M}}\right)_{T}\approx\frac{\overline{F_{s^*}}+F_{s^*,eddy}}{\overline{F_M}+F_{M,eddy}}\ . \tag{21}$$

Fig. 9 shows the scattered plot of $\left(\frac{\partial \overline{s^*}}{\partial \bar{M}}\right)_{\psi}$ against $\left(\frac{\partial \overline{s^*}}{\partial \bar{M}}\right)_{T}$ at $t_{RI}$ of all experiments near the outflow region. All data points fall near the diagonal line, with a mean value around $-5.5\times 10^{-5}\ [K^{-1}s^{-1}]$. The good agreement between the two variables confirms that $\left(\frac{\partial s^*}{\partial M}\right)_{T}$ near tropopause is controlled by the ratio of total entropy forcing to total momentum forcing at the outflow region. We note here that while the strong mixing near the outflow determines the precise value of $\left(\frac{\partial \overline{s^*}}{\partial \bar{M}}\right)_{T}$, the root cause of this mixing stems from the development of an inertially unstable region near the outflow region, as explained previous paragraph.

Peng et al. (2019) show that as the eyewall convective updrafts reach the tropopause, the upward transport of large values of $M$ and $s^*$ causes a bias of the parcels to develop into outflow, together with internal gravity wave generation and propagation. Our results presented herein similarly indicate that the development of transient eddies (deviation from 3-hourly time mean), which include internal gravity waves, can contribute to the mixing of $s^*$and $M$. Together with subgrid-scale turbulent mixing, they modulate the value of $\left(\frac{\partial \overline{s^*}}{\partial \overline{M}}\right)_T$ near the tropopause, constraining the structure of $M$ surfaces at the upper-level outflow region.

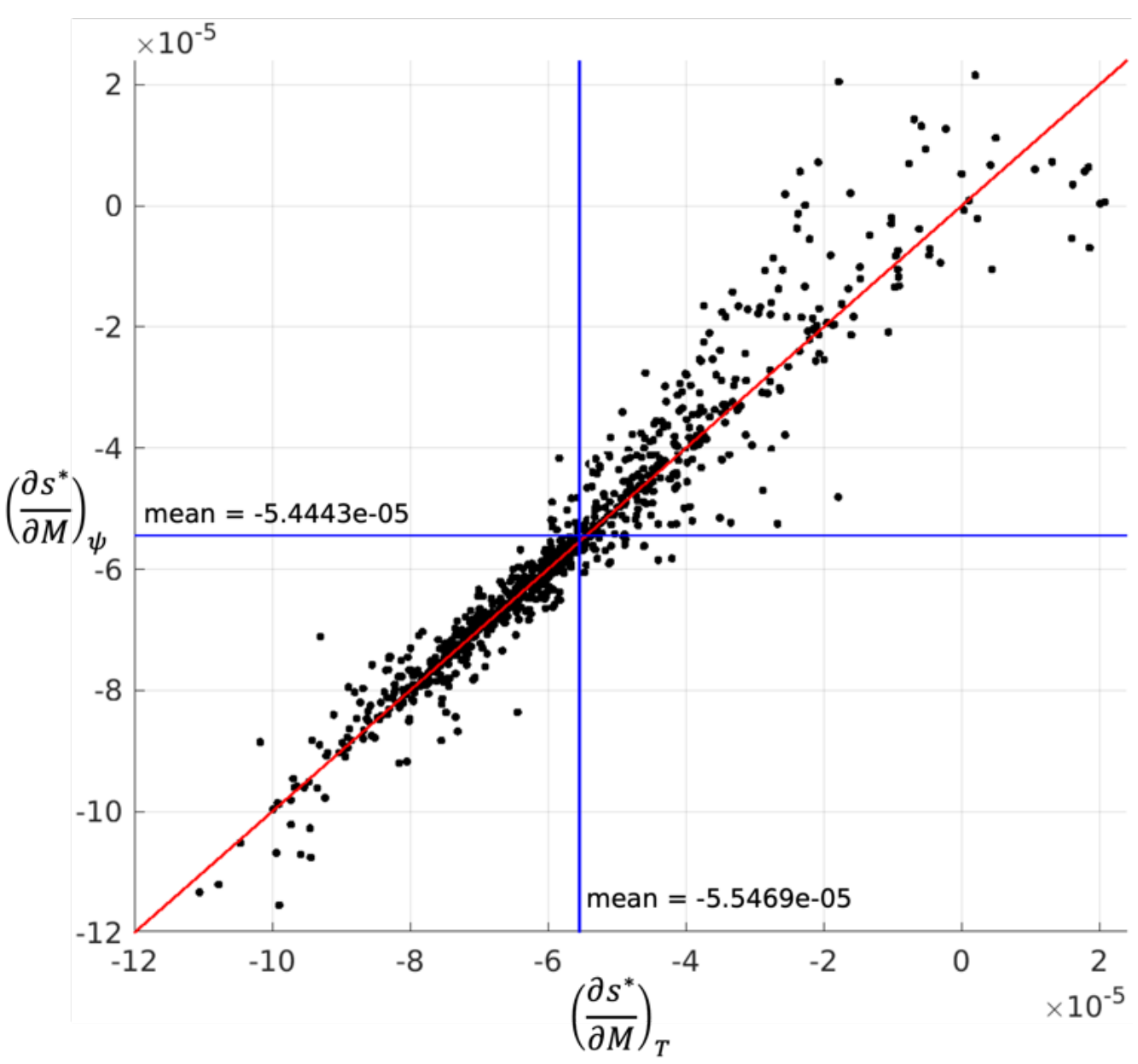

**Fig.9:** Scattered plot of $\left(\frac{\partial s^*}{\partial M}\right)_\psi$ versus $\left(\frac{\partial s^*}{\partial M}\right)_T$ of all members at $t_{RI}$. The data points are collected near the outflow region, with r between 30 and 80 km radii and z between 13.5 and 15 km. We exclude grid points where the normalized Jacobians $\frac{\left|J\left(\frac{M,T}{r,z}\right)\right|}{|\nabla M||\nabla T|}$ or $\frac{\left|J\left(\frac{M,\psi}{r,z}\right)\right|}{|\nabla M||\nabla \psi|}$ are less than 0.01 to avoid data near singularities of $\left(\frac{\partial s^*}{\partial M}\right)_\psi$ and $\left(\frac{\partial s^*}{\partial M}\right)_T$.

## 5. Discussion

*On the computation of $\frac{\partial s^*}{\partial M}$ and associated uncertainty*

Equation (8) demonstrates that, for a non-SN vortex, the slope of the inward bending $M$ surfaces of a balanced vortex is constrained by the *gradient between $s^*$ and $M$ along constant temperature*, i.e., $\left(\frac{\partial s^*}{\partial M}\right)_T$. A crucial point to highlight is that if the direction in which the gradient of $s^*$ is evaluated is not specified, $\frac{\partial s^*}{\partial M}$ is *not mathematically defined* and its magnitude (i.e., $\left|\frac{\partial s^*}{\partial M}\right|$) is *highly variable and uncertain*. As illustrated in Fig. 10a, if $\frac{\partial s^*}{\partial M}$ is evaluated along constant $s^*$

direction, then $\left|\left(\frac{\partial s^*}{\partial M}\right)_{s^*}\right|$ is *identically zero*. On the other hand, if $\frac{\partial s^*}{\partial M}$ is computed along constant $M$ direction, then $\left|\left(\frac{\partial s^*}{\partial M}\right)_{M}\right|$ is $+\infty$ (Fig. 10b)[4]. This strong sensitivity means that, for a non-SN vortex, $\left|\frac{\partial s^*}{\partial M}\right|$ can vary between 0 and $+\infty$, depending on slight variations in the direction of the derivative. This unconstrained freedom highlights the critical need to develop a rigorous framework to demonstrate how $\frac{\partial s^*}{\partial M}$ should be computed and linked to the balanced intensity under non-SN condition.

Importantly, our results presented herein proved that among all possible directions, it is the gradient along *constant temperature* that imposes physical constraints on the structure of a balanced vortex (Fig. 10c).

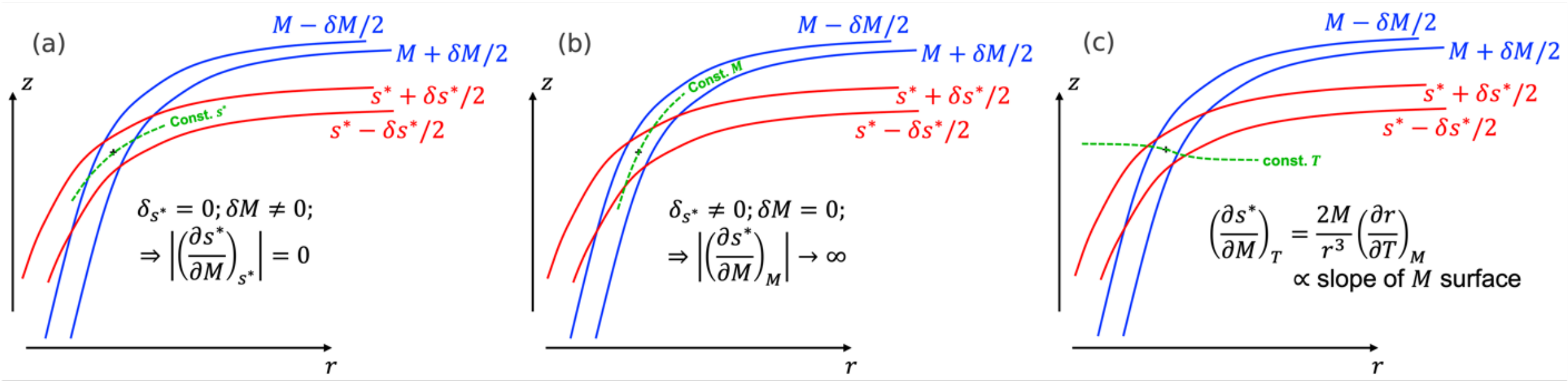

**Fig. 10:** Schematics illustrating the differences of partial derivative of $s^*$ w.r.t $M$ along different directions: (a) along $s^*$ isoline; (b) along $M$ isoline; and (c) along $T$ isoline. The direction where the derivative is held constant is shown by the green dashed line.

## 6. Conclusion

By revisiting the derivation of the potential intensity theory (PI, Emanuel 1986, Emanuel and Rotunno 2011, Bryan and Rotunno 2009a), this study relaxed the symmetric neutrality (SN) assumption and rigorously derived a generalized tangential wind formula that summarizes the dynamic and thermodynamic constraints on the vortex structure and intensity at the low level, revealing how previous formulations under SN condition (E86, BR09b) may be extended into non-symmetrically neutral (non-SN) regime.

Several important findings are identified by relaxing the SN assumption. First, it has been shown that in the saturated eyewall and upper-level outflow, the curvature of an absolute angular momentum $M$ isoline is dynamically tied to the *local derivative of the saturation entropy* $s^*$ *and* $M$ *holding temperature* $T$ *constant*, i.e., $\left(\frac{\partial s^*}{\partial M}\right)_T$, a generalization of Emanuel (1986). This constraint is shown to be sufficient to determine the balanced intensity at the low level, i.e., $\frac{v_b}{r_b} = \int_{T_b}^{T_o} \left(\frac{\partial s^*}{\partial M}\right)_T dT$. By holding $T_b$ and $T_o$ as constants (denoted as $T_b'$ and $T_o'$), this wind speed formula can be rewritten as $\frac{v_b}{r_b} = \frac{d}{dM}\left(\int_{T_b'}^{T_o'} s^* \, dT\right)$, which suggests that a TC eyewall is a circular frontal zone of integrated saturation entropy under non-SN conditions, a generalization of Emanuel (1997).

[4] Except when under SN condition where $s^*$ and $M$ are congruent, both $\left(\frac{\partial s^*}{\partial M}\right)_{s^*}$ and $\left(\frac{\partial s^*}{\partial M}\right)_{M}$ are in a indeterminate form (i.e., $\frac{0}{0}$), whose limits equal $\frac{ds^*}{dM}$.

By incorporating the radial and vertical momentum advection and momentum forcings, we derived a generalized unbalanced term that converges to the supergradient wind component derived by Bryan and Rotunno (2009b) under the SN assumption. The relaxation of the SN assumption also allows us to obtain additional terms that represent contributions from radial and vertical momentum forcing within the TC boundary layer. These terms combined to form the generalized $v_{max}$ formula that applies within the TC boundary layer before the SN assumption is valid.

An ensemble of axisymmetric numerical simulations was conducted to verify the validity of the generalized $v_{max}$ formula. The results showed that this generalized formula can accurately quantify the contributions of the balanced, unbalanced and frictional wind components during the intensification period, demonstrating that it is the proper extension of the traditional PI (balanced and/or unbalanced) $v_{max}$ formula into the non-SN regime.

Focusing on the 24-hour period centered at the time of the maximum intensification rate, several interesting findings regarding the balanced component are revealed. Before the intensification rate reaches peak value, $\left(\frac{\partial s^*}{\partial M}\right)_T$ within the saturated eyewall first develops a top-heavy, linear structure as a function of temperature, with large negative value occurring at the upper troposphere near temperature of 210-240 K. This linear dependence on temperature confirms that the SN assumption is not valid at the time of peak intensification rate. The top-heavy structure of $\left(\frac{\partial s^*}{\partial M}\right)_T$ during the early phase of intensification suggests that the development of a deep tall vortex preconditions the onset of intensification process, consistent with recent observational evidence (Fischer et al. 2022, 2025; DesRosiers, et al. 2023). Further examination of the outflow region shows that the large negative $\left(\frac{\partial s^*}{\partial M}\right)_T$ value at the upper level is amplified by the small inertial stability at the outflow region, which is shown to be dynamically linked to the presence of negative moist potential vorticity (PV) at the lower troposphere using the conservation principle of moist PV. The precise value of the large negative $\left(\frac{\partial s^*}{\partial M}\right)_T$ at the upper level is controlled by the ratio of the entropy forcing to the momentum forcing (mixing due to transient eddies) at the outflow region. These results indicate the importance of resolved and parameterized mixing at the upper level in controlling the inward bending structure of $M$ surfaces at the tropopause outflow region, consistent with Emanuel (2012).

The theoretical results identified in the study have other important implications. First, the generalized balanced $v_{max}$ component shows that the distribution of saturation entropy $s^*$ along the $M$ surfaces at the vicinity of eyewall updraft sufficiently determines the balanced intensity at the low level (equations 9 and 10), providing an explicit linkage for investigating the impacts of ventilation on TC intensity under non-SN conditions. Second, analysis of the simulations suggested that tropopause mixing determines the overall vertical variation of $\left(\frac{\partial s^*}{\partial M}\right)_T$ in the free troposphere, which modulates the magnitude and potential roles of vertical advection of entropy in the intensification process, an effect neglected by previous studies due to the SN assumption.

Because the goal of this study is to derive and explore the detailed vortex dynamics in the non-SN regime, many aspects of the results cannot be covered due to the length of the manuscript. These include examining the sensitivity of our findings on the various experimental parameters, such as horizontal mixing length, and the ratio of drag and enthalpy exchange coefficients. A future study will be dedicated to developing a time-dependent intensification theory that captures non-SN vortex structure revealed in the present study.

**Acknowledgments**

I thank Drs. Richard Rotunno, Kerry Emanuel, Daniel Chavas, Brian Tang, Robert Fovell, Bolei Yang, Chris Davis and Zheng Wu for their insightful discussion and encouragement. I also greatly appreciate the three anonymous reviewers for the careful and detailed reviews, which provided immensely valuable feedback to improve the clarity and correctness of the major findings of this work.

**Data availability statement**

The CM1 namelist files and model output are available upon request via a University at Albany Thematic Real-time Environmental Distributed Data Services (THREDDS) server.

## Appendix A: Five equivalent forms of a Jacobian Matrix Determinant

This appendix derives a few equivalent Jacobian forms used to obtain some of the major findings of this study.

Let $J$ be the Jacobian matrix determinant of a two-dimensional transformation from $(M, p)$ space to $(s^*, T)$ space, defined as

$$J\left(\frac{s^*, T}{M, p}\right) \coloneqq \det\left(\begin{bmatrix} \frac{\partial s^*}{\partial M} & \frac{\partial s^*}{\partial p} \\ \frac{\partial T}{\partial M} & \frac{\partial T}{\partial p} \end{bmatrix}\right) ,$$

$$= \left(\frac{\partial T}{\partial p}\right)_M \left(\frac{\partial s^*}{\partial M}\right)_p - \left(\frac{\partial T}{\partial M}\right)_p \left(\frac{\partial s^*}{\partial p}\right)_M . \tag{A1}$$

If we factor out $\left(\frac{\partial s^*}{\partial M}\right)_p$ in (A.1), we can then use the triple product rule of partial derivatives $\left(\frac{\partial s^*}{\partial p}\right)_M \left(\frac{\partial M}{\partial s^*}\right)_p \left(\frac{\partial p}{\partial M}\right)_{s^*} = -1$ and the fact that $\left(\frac{\partial T}{\partial p}\right)_{s^*} = \left(\frac{\partial T}{\partial p}\right)_M + \left(\frac{\partial T}{\partial M}\right)_p \left(\frac{\partial M}{\partial p}\right)_{s^*}$ to rewrite $J\left(\frac{s^*,T}{M,p}\right)$

$$J\left(\frac{s^*, T}{M, p}\right) = \left(\frac{\partial s^*}{\partial M}\right)_p \left(\left(\frac{\partial T}{\partial p}\right)_M + \left(\frac{\partial T}{\partial M}\right)_p \left(\frac{\partial M}{\partial p}\right)_{s^*}\right) = \left(\frac{\partial s^*}{\partial M}\right)_p \left(\frac{\partial T}{\partial p}\right)_{s^*} . \tag{A2}$$

Following the same procedure and factoring the other three partial derivatives of (A.1), we show that $J\left(\frac{s^*,T}{M,p}\right)$ may be written in at least four alternative equivalent ways:

$$J\left(\frac{s^*, T}{M, p}\right) = \left(\frac{\partial T}{\partial p}\right)_{s^*} \left(\frac{\partial s^*}{\partial M}\right)_p , \tag{A3a}$$

$$= \left(\frac{\partial T}{\partial p}\right)_M \left(\frac{\partial s^*}{\partial M}\right)_T , \tag{A3b}$$

$$= -\left(\frac{\partial s^*}{\partial p}\right)_M \left(\frac{\partial T}{\partial M}\right)_{s^*} , \tag{A3c}$$

$$= -\left(\frac{\partial T}{\partial M}\right)_p \left(\frac{\partial s^*}{\partial p}\right)_T . \tag{A3d}$$

Each of (A.3a)-(A.3d) represents a specific order of transformation from $(M, p)$ to $(T, s)$ spaces, as shown in Figure A1; while (A.1) only specifies the start- and end-point of the transformation. Note that the Jacobian for transformation from $(s^*, T)$ to $(T, s^*)$ is -1, meaning that the dimensional order of the coordinate system matters (Fig. A1b).

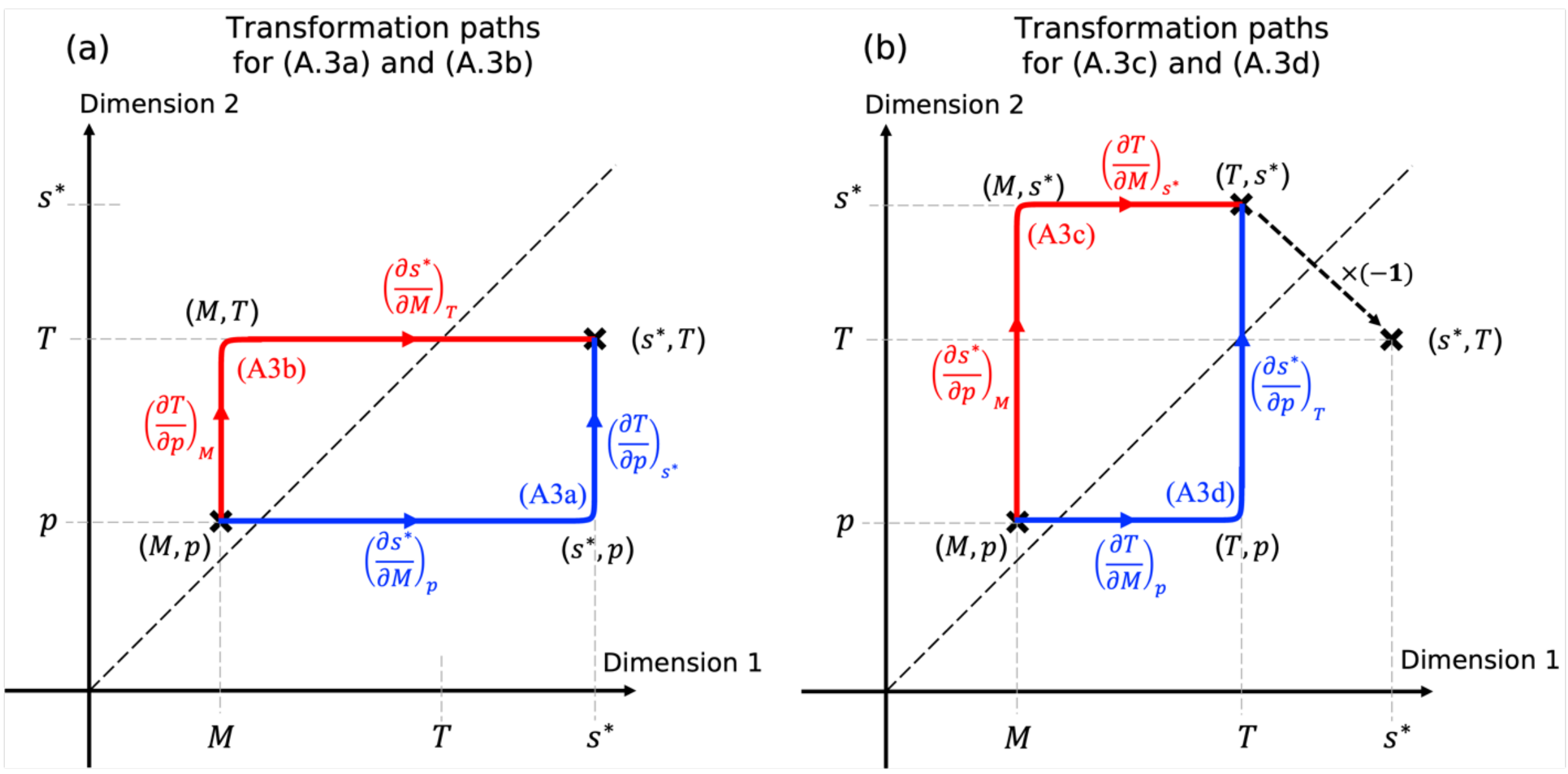

**Fig.A1:** (a) A schematics illustrating the transformation paths from $(M,p)$ space to $(s^*,T)$ space based on equations A.3a (blue) and A.3b (red). The Jacobians determinant for each transformation step is shown next to the transformation path. The horizontal and vertical axes represent all possible coordinate variables; each point in the two-dimensional space is a unique coordinate system. (b) is similar to (a), but for equations A.3c (red) and A.3d (blue). The thick black dashed arrow represents the transformation from $(T,s^*)$ to $(s^*,T)$ space, which has a Jacobian determinant of $-1$.

## Appendix B: Derivation of balanced tangential wind using the energetic balance (5)

The balanced tangential wind shown in equation (9) was derived using one of the Jacobian forms (A.3b). In this appendix, we derive (9) and (10) based on the energetic property (5), i.e., $dM^2 d\left(-\frac{1}{2r^2}\right) = ds^* dT$.

For a given $M$ surface ($M_0$) within the saturated eyewall region, without loss of generality, we consider the narrow region bounded by the two adjacent $M$ surfaces surrounding $M_0$ (i.e., $M_{-1} = M_0 - \frac{\delta M}{2}$ and $M_1 = M_0 + \frac{\delta M}{2}$, with infinitesimal $\delta M$) and the two *constant* temperature surfaces $T_b$ and $T_o$, as shown in Fig. B1a. Here, $T_o$ is the temperature at the outflow radius $r_o$ where $v$ at the $M_0$ surface becomes 0. Because $M$ is the vertical axis in $(-1/2r^2, M^2)$ space and $\delta M$ is infinitesimal, the region of interest in $(-1/2r^2, M^2)$ space is an infinitesimally shallow trapezoid (Fig. B1b), with thickness $M_1^2 - M_{-1}^2 \approx 2M_0\delta M$. To the first order of $\delta M$, the area of this region can be computed as a shallow parallelogram of width $\left(-\frac{1}{2r_o^2} + \frac{1}{2r_b^2}\right)$.

$$Area_{(M^2,-1/2r^2)} = 2M_0\delta M \times \left(-\frac{1}{2r_o^2} + \frac{1}{2r_b^2}\right). \quad \text{(B1)}$$

Because $M_0 = \frac{1}{2}fr_o^2 = r_b v_b + \frac{1}{2}fr_b^2$, $Area_{(M^2,-1/2r^2)} = \delta M \times \left(-\frac{2M_0}{2r_o^2} + \frac{2M_0}{2r_b^2}\right) = \frac{v_b}{r_b}\delta M$.

On the other hand, the structure of the same region in $(s^*, T)$ space depends on the shape of the $M_{-1}$ and $M_1$ isolines in $(s^*, T)$ space, as shown in Fig. B1c. To calculate the area, the first approach would be to discretize the region into thin layers with infinitesimal (negative) $\delta T$. The width $\delta s^*$ of each layer can be computed as $\left(\frac{\partial s^*}{\partial M}\right)_T \delta M$. Therefore, $Area_{(s^*,T)}$ can be computed as

$$Area_{(s^*,T)} = \sum_{T=T_b}^{T_o} \left(\frac{\partial s^*}{\partial M}\right)_T \delta M \delta T = \delta M \sum_{T=T_b}^{T_o} \left(\frac{\partial s^*}{\partial M}\right)_T \delta T \ . \quad \text{(B2)}$$

Taking $\delta T \to 0$, $\delta M \sum_i \left(\frac{\partial s^*}{\partial M}\right)_T \delta T$ becomes $\delta M \int_{T_b}^{T_o} \left(\frac{\partial s^*}{\partial M}\right)_T dT$. By the energetic property (5), we have $\frac{v_b}{r_b}\delta M = \delta M \int_{T_b}^{T_o} \left(\frac{\partial s^*}{\partial M}\right)_T dT$. Canceling $\delta M$ yields

$$\frac{v_b}{r_b} = \int_{T_b}^{T_o} \left(\frac{\partial s^*}{\partial M}\right)_T dT\ . \quad \text{(B3)}$$

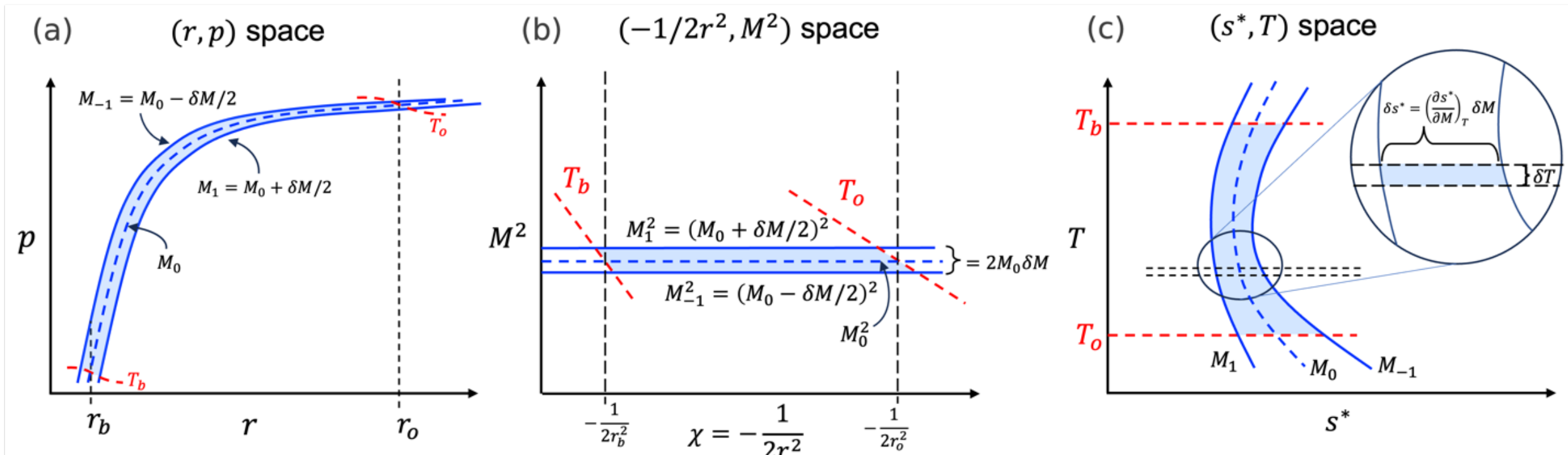

**Fig. B1:** Schematics illustrating the energy equipartition principle (equation 5): (a) region of interest centered along a given $M_0$ surface, bounded left and right by two adjacent $M$ surfaces ($M_{-1}$ and $M_1$, separated by an infinitesimal $\delta M$) and above and below by $T_b$ and $T_o$; (b) the corresponding region in $(-1/2r^2, M^2)$ space; (c) the corresponding region in $(s^*, T)$ space. Note that the y-axis in (a) is pointing upward to indicate the direction where height increases.

An alternative approach to calculate the $Area_{(s^*,T)}$ is to define an area function $A(M) = \int_{T_b}^{T_o} s^* dT$, which is the *negative area to the left of the M isoline between* $T_o$ *and* $T_b$ (Fig. B2a,b). The area of our interest (Fig. B1c) is $A(M_1) - A(M_{-1}) = \delta\left(\int_{T_b}^{T_o} s^* dT\right)$, i.e., the change in saturation entropy integral between the two $M$ surfaces. To the first order of $\delta M$,

$$
\begin{aligned}
Area_{(s^*,T)} = \delta\left(\int_{T_b}^{T_o} s^* dT\right) &= A(M_1) - A(M_{-1}) \ , \\
&\approx \frac{d(A)}{dM}\delta M \ , \\
&= \frac{d}{dM}\left(\int_{T_b}^{T_o} s^* dT\right)\delta M \ . \qquad \text{(B4)}
\end{aligned}
$$

Note that both $A(M_1)$ and $A(M_{-1})$ are negative and that $A(M_1) > A(M_{-1})$, resulting in positive area $\frac{d}{dM}\left(\int_{T_b}^{T_o} s^* dT\right)\delta M$ in (B4). Equating $Area_{(M^2,-1/2r^2)}$ and $Area_{(s^*,T)}$, we obtain

$$
\frac{v_b}{r_b} = \frac{d}{dM}\left(\int_{T_b}^{T_o} s^* dT\right) \ . \qquad \text{(B5)}
$$

The above two approaches are mathematically linked by the Leibniz rule of integration. We verify their equivalency by plotting $v_b$ computed from (B.3) against (B.5) for each of the 11 members of the CTRL at $t_{RI}$, as shown in Fig. B2d.

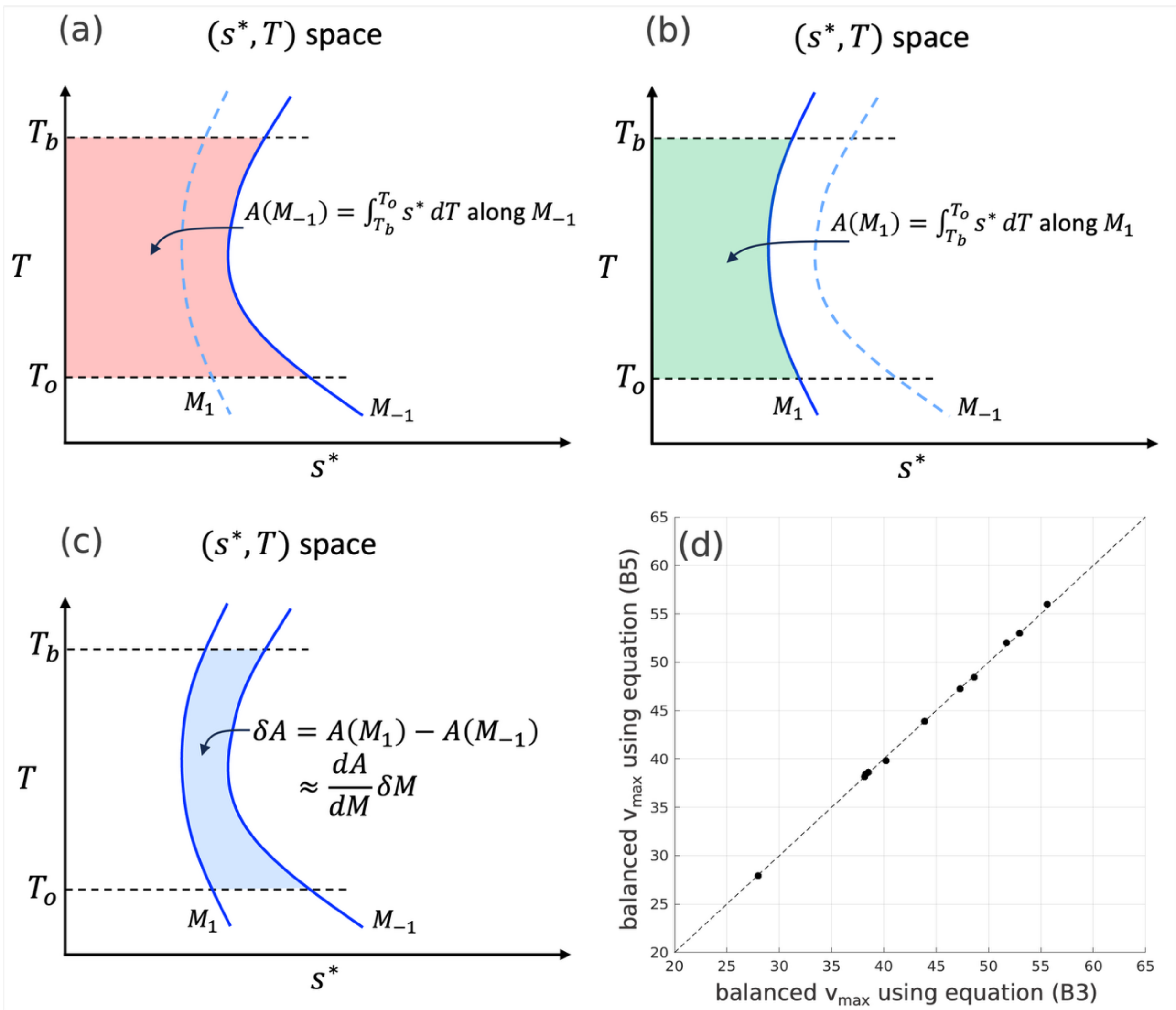

**Fig. B2:** Schematics illustrating a second method of evaluating the area of interest in $(s^*, T)$ space: (a) the area to the left of the $M_{-1}$ isoline in $(s^*, T)$ space (bounded by $T_b$ and $T_o$), which equals to $A(M_{-1})$; (b) similar to (a), but for area represented by $A(M_1)$; (c) the area of interested in $(s^*, T)$ space, which equals to the difference between $A(M_1)$ and $A(M_{-1})$. (d) A scattered plot showing the balanced $v_b$ evaluated from (B.5) against that from (B.3) for each CTRL member at $t_{RI}$.

## Appendix C: The generalized tangential wind formula

In this appendix, we derive the fully generalized tangential wind formula in height coordinate, which allows us to compare with results from Lilly 1979, 1986 (unpublished manuscripts, Tao et al. 2020) and BR09b.

The radial, vertical momentum equations and anelastic form of the continuity in $(r, z)$ coordinate at issue are

$$\frac{\partial u}{\partial t} + u\frac{\partial u}{\partial r} + w\frac{\partial u}{\partial z} = \frac{M^2}{r^3} - \frac{1}{4}f^2 r - \alpha\frac{\partial p}{\partial r} + F_r\,, \tag{C1a}$$

$$\frac{\partial w}{\partial t} + u\frac{\partial w}{\partial r} + w\frac{\partial w}{\partial z} = -g - \alpha\frac{\partial p}{\partial z} + F_z\,, \tag{C1b}$$

$$\frac{1}{r}\frac{\partial r\rho u}{\partial r} + \frac{\partial \rho w}{\partial z} = 0\,, \tag{C1c}$$

where $\rho$, $\alpha$ and $p$ are the total moist air density, specific volume and pressure; $F_r$ and $F_z$ are the momentum forcings in the radial and vertical directions. For simplicity, we neglect the hydrometeor loading in the moist air density $\rho = \rho_d + \rho_v$, where $\rho_v$ is vapor density. With equation (C1c), $\boldsymbol{u} = (u, w)$ can be written in terms of nondivergent streamfunction $\psi$ as $\left(\frac{1}{r\rho}\frac{\partial\psi}{\partial z}, \frac{-1}{r\rho}\frac{\partial\psi}{\partial r}\right)$. Following Lilly 1979, (C1a) and (C1b) can then be rewritten as

$$\frac{\partial u}{\partial t} + \frac{\partial}{\partial r}\left(\frac{u^2 + w^2 + v^2}{2} + \frac{fM}{2} + gz\right) - \frac{\eta}{r\rho}\frac{\partial\psi}{\partial r} = \frac{1}{2r^2}\frac{\partial M^2}{\partial r} - \alpha\frac{\partial p}{\partial r} + F_r\,, \tag{C2a}$$

$$\frac{\partial w}{\partial t} + \frac{\partial}{\partial z}\left(\frac{u^2 + w^2 + v^2}{2} + \frac{fM}{2} + gz\right) - \frac{\eta}{r\rho}\frac{\partial\psi}{\partial z} = \frac{1}{2r^2}\frac{\partial M^2}{\partial z} - \alpha\frac{\partial p}{\partial z} + F_z\,, \tag{C2b}$$

where $\eta = \left(\frac{\partial u}{\partial z} - \frac{\partial w}{\partial r}\right)$ is the azimuthal vorticity. Defining $\tilde{\eta} = \eta/\rho$ and cross differentiating (C2a) and (C2b) yields

$$\frac{\partial\eta}{\partial t} = \frac{2M}{r^3}\frac{\partial M}{\partial z} + J\left(\frac{\alpha, p}{r, z}\right) + J\left(\frac{\psi, \tilde{\eta}/r}{r, z}\right) + \frac{\partial F_r}{\partial z} - \frac{\partial F_z}{\partial r}\,. \tag{C3}$$

We note here that $\frac{2M}{r^3}\frac{\partial M}{\partial z} = \nabla \times \left(\left(\frac{v^2}{r} + fv\right)\hat{r}\right) \cdot \hat{\lambda}$ is centrifugal torque; $J\left(\frac{\alpha, p}{r, z}\right) = \nabla \times (-\alpha\nabla p) \cdot \hat{\lambda}$ is baroclinic torque in azimuthal direction; and $J\left(\frac{\psi, \tilde{\eta}/r}{r, z}\right) = -\frac{\partial}{\partial r}\left(\frac{\eta}{r\rho}\frac{\partial\psi}{\partial z}\right) - \frac{\partial}{\partial z}\left(-\frac{\eta}{r\rho}\frac{\partial\psi}{\partial r}\right) = -\nabla \cdot (\eta\boldsymbol{u})$ is the convergence of azimuthal vorticity flux. (C3) thus quantifies how the imbalance between the centrifugal torque, the baroclinic torque, azimuthal vorticity flux convergence and frictional torque drives the azimuthal vorticity tendency and the associated acceleration of the secondary circulation.

To couple with saturated moist dynamics, we assume saturation at the eyewall updraft region and define $s^* = c_p \ln T - R_d \ln p_d + \frac{L_v q^*}{T}$, which satisfies the modified first law of thermodynamics (E86, BR09b)

$$T ds^* = c_p dT - \alpha_d dp + d(L_v q^*) \ . \tag{C4}$$

Integrating (C4) along the closed boundary $\partial D$ of an arbitrary saturated region $D$ within the eyewall and applying the Stoke's theorem, we have

$$\left(\nabla T \times \nabla s^* \cdot \hat{\lambda}\right) = \left(\nabla p \times \nabla \alpha_d \cdot \hat{\lambda}\right) \ ,$$

$$J\left(\frac{s^*, T}{r, z}\right) = J\left(\frac{\alpha_d, p}{r, z}\right), \tag{C5}$$

where $\hat{\lambda}$ is the unit vector in the azimuthal direction. Additionally, we neglect the contributions of water vapor to the value of density, such that $J\left(\frac{\alpha_d, p}{r,z}\right) = J\left(\frac{\alpha, p}{r,z}\right)$. Therefore, (C5) becomes

$$J\left(\frac{s^*, T}{r, z}\right) = J\left(\frac{\alpha, p}{r, z}\right) \ . \tag{C6}$$

We will further neglect $\frac{\partial \eta}{\partial t}$, which is typically one to two orders smaller than the rest of the terms in (C3). Combining (C6) with (C3), using equation (A3) to write $J\left(\frac{\psi, \tilde{\eta}/r}{r,z}\right) = \left(\frac{\partial \psi}{\partial r}\right)_z \left(\frac{\partial \tilde{\eta}/r}{\partial z}\right)_\psi$, $J\left(\frac{s^*, T}{r,z}\right) = \left(\frac{\partial T}{\partial z}\right)_{s^*} \left(\frac{\partial s^*}{\partial r}\right)_z$ and dividing both sides of (C3) with $\left(\frac{\partial M}{\partial r}\right)_z$, we have

$$\frac{2M}{r^3}\left(\frac{\partial r}{\partial z}\right)_M = \left(\frac{\partial T}{\partial z}\right)_{s^*}\left(\frac{\partial s^*}{\partial M}\right)_z + \left(\frac{\partial \psi}{\partial M}\right)_z\left(\frac{\partial \tilde{\eta}/r}{\partial z}\right)_\psi + \left(\frac{\partial F_r}{\partial z}\right)_r\left(\frac{\partial r}{\partial M}\right)_z - \left(\frac{\partial F_z}{\partial M}\right)_z . \tag{C7}$$

Using the equivalent forms of Jacobian determinant (A3), (C7) can be further rewritten as

$$\frac{2M}{r^3}\left(\frac{\partial r}{\partial z}\right)_M = \left(\frac{\partial T}{\partial z}\right)_M\left(\frac{\partial s^*}{\partial M}\right)_T + \left(\frac{\partial \psi}{\partial M}\right)_{\tilde{\eta}/r}\left(\frac{\partial \tilde{\eta}/r}{\partial z}\right)_M - \left(\frac{\partial F_r}{\partial M}\right)_r\left(\frac{\partial r}{\partial z}\right)_M - \left(\frac{\partial F_z}{\partial M}\right)_z . \tag{C8}$$

Integrating (C8) along an $M$ isoline from $z_b$ to the outflow radius at outflow level $z_o$, we obtain the generalized tangential wind formula in height coordinate

$$\frac{v_b}{r_b} = \int_{T_b}^{T_o}\left(\frac{\partial s^*}{\partial M}\right)_T dT + \int_{\tilde{\eta}_b/r_b}^{\tilde{\eta}_o/r_o}\left(\frac{\partial \psi}{\partial M}\right)_{\tilde{\eta}/r} d\left(\frac{\tilde{\eta}}{r}\right) - \int_{r_b}^{r_o}\left(\frac{\partial F_r}{\partial M}\right)_r dr - \int_{z_b}^{z_o}\left(\frac{\partial F_z}{\partial M}\right)_z dz \ . \tag{C9}$$

**Appendix D: Correction term to the balanced wind component**

In the derivation of Eq. 9 and C6, we neglected the contribution of water vapor density to the balanced intensity. In this appendix, we derive a correction term to account for the effect of this approximation on the balanced wind component.

As derived in Appendix C, $s^* = c_p \ln T - R_d \ln p_d + \frac{L_v q^*}{T}$ satisfies the following differential form $T ds^* = d(c_p T) - \alpha_d dp + d(L_v q^*)$ and Jacobian relation $J\left(\frac{s^*,T}{r,z}\right) = J\left(\frac{\alpha_d,p}{r,z}\right)$. To include the effect of vapor density in the balanced intensity, we first write $\rho = \rho_d + \rho_v^*(T)$, so that $\alpha = (\rho_d + \rho_v^*)^{-1} \approx \alpha_d + \alpha'$ with $\alpha' := -\frac{\rho_v^*}{\rho_d}\alpha_d$. Then, by expanding $J\left(\frac{\alpha,p}{r,z}\right) = \left(\nabla p \times \nabla \alpha \cdot \hat{\lambda}\right)$ as $\left(\nabla p \times \nabla \alpha_d \cdot \hat{\lambda}\right) + \left(\nabla p \times \nabla \alpha' \cdot \hat{\lambda}\right)$ and using the Clausius–Clapeyron equation $\frac{de_s}{dT} = \frac{L_v e_s}{R_v T^2}$, we can show that, to the first order of $q^* \left(= \frac{\rho_v^*}{\rho_d} \ll 1\right)$, $J\left(\frac{\alpha,p}{r,z}\right)$ may rewrite as

$$\left(\nabla p \times \nabla \alpha \cdot \hat{\lambda}\right) \approx \left(\nabla p \times \nabla \alpha_d \cdot \hat{\lambda}\right)(1-\gamma)\,, \qquad \text{(D1)}$$

where $\gamma = \left(\frac{1+\frac{L_v}{R_v T}}{1+\frac{L_v q^*}{R_d T}}\right) q^*$. This correction is similar to the unapproximated form derived in Emanuel (1988), but does not require the derivation of Maxwell relation and does not assume the total water content being a function of $M$ alone. Because $J\left(\frac{s^*,T}{r,z}\right) = J\left(\frac{\alpha_d,p}{r,z}\right)$ (from Eq. C5), D1 can be rewritten as

$$J\left(\frac{\alpha,p}{r,z}\right) = J\left(\frac{s^*,T}{r,z}\right)(1-\gamma)\,. \qquad \text{(D2)}$$

This relation is the bias-corrected version of (C6) in appendix C. Using (D2) and continuing the derivation in appendix C, we can obtain the following version of generalized tangential wind formula

$$\frac{v_b}{r_b} = \int_{T_b}^{T_o} \left(\frac{\partial s^*}{\partial M}\right)_T dT + \int_{\tilde{\eta}_b/r_b}^{\tilde{\eta}_o/r_o} \left(\frac{\partial \psi}{\partial M}\right)_{\tilde{\eta}/r} d\left(\frac{\tilde{\eta}}{r}\right) - \int_{r_b}^{r_o} \left(\frac{\partial F_r}{\partial M}\right)_r dr - \int_{z_b}^{z_o} \left(\frac{\partial F_z}{\partial M}\right)_z dz - \int_{T_b}^{T_o} \gamma \left(\frac{\partial s^*}{\partial M}\right)_T dT. \qquad \text{(D6)}$$

The last term $-\int_{T_b}^{T_o} \gamma \left(\frac{\partial s^*}{\partial M}\right)_T dT$ is the correction term to account for the impact of the vapor density on the balanced intensity. This correction term is strictly true under saturation condition.

## Appendix E: Regularization of singularity in $\left(\frac{\partial s^*}{\partial M}\right)_T$ and $\left(\frac{\partial \psi}{\partial M}\right)_{\tilde{\eta}}$

In this appendix, we document the treatment to singularities of $\left(\frac{\partial s^*}{\partial M}\right)_T$ and $\left(\frac{\partial \psi}{\partial M}\right)_{\tilde{\eta}}$ along the integration path when they occur. The following discussion will only focus on $\left(\frac{\partial s^*}{\partial M}\right)_T$ for simplicity, since the same reasoning and steps can also be applied to analyze the $\left(\frac{\partial \psi}{\partial M}\right)_{\tilde{\eta}}$ term.

From the definition $\left(\frac{\partial s^*}{\partial M}\right)_T \coloneqq \lim_{\substack{\delta M \to 0 \\ T=const}} \frac{\delta s^*}{\delta M}$, we can see that the necessary and sufficient condition for the occurrence of singularity is when $\delta M$ becomes *identically zero* when holding $T$ constant, which happens if and only if the isolines of $M$ and $T$ are parallel at a grid point. However, since our interest is the integral $\int \left(\frac{\partial s^*}{\partial M}\right)_T dT$ , I will show that $\int \left(\frac{\partial s^*}{\partial M}\right)_T dT$ remains integrable even if $\left(\frac{\partial s^*}{\partial M}\right)_T$ may have (a finite number of) singularities along the integration path. First, we note that at the singularity where $M$ and $T$ isolines are parallel, $dT$ must also be 0. As such, the integrand $\left(\frac{\partial s^*}{\partial M}\right)_T dT$ must be in an indeterminant form of $\infty \times 0$ at the point of singularity. This means we can regularize the singularity if this indeterminant form has a finite limit.

To show this, we will first change the variable of integration to something that always has non-zero finite differential along the integration path. The natural choice of such a variable is the path length, denoted as $l$. First, we construct the natural coordinate along the integration path (an $M$ isoline), denoted by the unit vectors $(\hat{n}, \hat{l})$. Specifically, we define $\hat{n} = \frac{\nabla M}{|\nabla M|}$ to be the unit vector perpendicular to the $M$ isoline, while $\hat{l} \coloneqq \hat{n} \times \hat{\lambda}$ is the direction along the $M$ isoline ($\hat{\lambda}$ is the azimuthal direction). Note that the two unit vectors $(\hat{n}, \hat{l})$ form a local cartecian coordinate $(n, l)$ of unit grid size in $(r, z)$ space. We now consider the integral of $\left(\frac{\partial s^*}{\partial M}\right)_T$ across the interval $\left(T_i - \frac{\delta T}{2}, T_i + \frac{\delta T}{2}\right)$ where the singularity lies. We can write this integral using the natural coordinate as

$$\int_{T_i-\frac{\delta T}{2}}^{T_i+\frac{\delta T}{2}} \left(\frac{\partial s^*}{\partial M}\right)_T dT = \int_{l_i-\frac{\delta l}{2}}^{l_i+\frac{\delta l}{2}} \left(\frac{\partial s^*}{\partial M}\right)_T \left(\frac{\partial T}{\partial l}\right)_M dl\,,$$

$$= \int_{l_i-\frac{\delta l}{2}}^{l_i+\frac{\delta l}{2}} \left(\frac{\partial s^*}{\partial M}\right)_l \left(\frac{\partial T}{\partial l}\right)_{s^*} dl\,,$$

$$= \int_{l_i-\frac{\delta l}{2}}^{l_i+\frac{\delta l}{2}} \left(\frac{\partial n}{\partial M}\right)_l \left(\frac{\partial s^*}{\partial n}\right)_l \left(\frac{\partial T}{\partial l}\right)_{s^*} dl\,, \tag{E1}$$

where $\left(l_i - \frac{\delta l}{2}, l_i + \frac{\delta l}{2}\right)$ is the incremental line segment of the integration path. The second equality uses the Jacobian identity (A.3) that $\left(\frac{\partial s^*}{\partial M}\right)_T \left(\frac{\partial T}{\partial l}\right)_M = \left(\frac{\partial s^*}{\partial M}\right)_l \left(\frac{\partial T}{\partial l}\right)_{s^*}$. Note that by definition $\left(\frac{\partial M}{\partial n}\right)_l \equiv |\nabla M|$ and that $\left(\frac{\partial s^*}{\partial n}\right)_l \left(\frac{\partial T}{\partial l}\right)_{s^*}$ is the Jacobian $J\left(\frac{s^*,T}{n,l}\right)$, which also equals to $J\left(\frac{s^*,T}{r,z}\right)$ because $J\left(\frac{n,l}{r,z}\right) = 1$ (i.e., $(n, l)$ is a local cartecian coordinate of unit grid size in $(r, z)$ space). Therefore, we can rewrite (E1) as

$$\int_{T_i-\frac{\delta T}{2}}^{T_i+\frac{\delta T}{2}} \left(\frac{\partial s^*}{\partial M}\right)_T dT = \int_{l_i-\frac{\delta l}{2}}^{l_i+\frac{\delta l}{2}} \frac{1}{|\nabla M|} J\left(\frac{s^*,T}{r,z}\right) dl\,. \tag{E2}$$

Within the vortex region, $dl$ and $J\left(\frac{s^*,T}{r,z}\right)$ are both finite and $|\nabla M|$ is always non-zero. This thus shows that $\left(\frac{\partial s^*}{\partial M}\right)_T dT$ indeed has a finite limit, which equals to $\frac{1}{|\nabla M|} J\left(\frac{s^*,T}{r,z}\right) dl$. We now define the regularization of $\left(\frac{\partial s^*}{\partial M}\right)_T$ (denoted as $\widetilde{\left(\frac{\partial s^*}{\partial M}\right)}_T$) as $\widetilde{\left(\frac{\partial s^*}{\partial M}\right)}_T \coloneqq \frac{1}{\delta T} \int_{T_i-\frac{\delta T}{2}}^{T_i+\frac{\delta T}{2}} \left(\frac{\partial s^*}{\partial M}\right)_T dT$. Given (E.2), we have

$$\widetilde{\left(\frac{\partial s^*}{\partial M}\right)}_T = \frac{1}{\delta T} \int_{l_i-\frac{\delta l}{2}}^{l_i+\frac{\delta l}{2}} J\left(\frac{s^*,T}{r,z}\right) \frac{dl}{|\nabla M|}\,. \tag{E3}$$

Close examination of (E.3) found (not shown) that when singularity does not occur, $\widetilde{\left(\frac{\partial s^*}{\partial M}\right)}_T$ equals to the mean value of $\left(\frac{\partial s^*}{\partial M}\right)_T$ over the interval $\left(T_i - \frac{\delta T}{2}, T_i + \frac{\delta T}{2}\right)$; while at intervals where singularity lies, it allows accurate integration across the singularity. Similarly, $\left(\frac{\partial \psi}{\partial M}\right)_{\tilde{\eta}}$ can be regularized as

$$\widetilde{\left(\frac{\partial \psi}{\partial M}\right)}_{\tilde{\eta}/r} \coloneqq \frac{1}{\delta(\tilde{\eta}/r)} \int_{l_i-\frac{\delta l}{2}}^{l_i+\frac{\delta l}{2}} J\left(\frac{\psi, \tilde{\eta}/r}{r,z}\right) \frac{dl}{|\nabla M|}\,. \tag{E4}$$

**References:**

Abarca, S. F., M. T. Montgomery, and J. C. McWilliams (2015), The azimuthally averaged boundary layer structure of a numerically simulated major hurricane, J. Adv. Model. Earth Syst..

Bister, M., and K. A. Emanuel, 1998: Dissipative heating and hurricane intensity. Meteor. Atmos. Phys., 65, 233–240.

Bryan, G. H., and J. M. Fritsch, 2002: A Benchmark Simulation for Moist Nonhydrostatic Numerical Models. Mon. Wea. Rev., 130, 2917–2928.

Bryan, G. H., 2008: On the computation of pseudoadiabatic entropy and equivalent potential temperature. Mon. Wea. Rev., 136, 5239–5245.

Bryan, G. H., and R. Rotunno, 2009a: The maximum intensity of tropical cyclones in axisymmetric numerical model simulations. Mon. Wea. Rev., 137, 1770–1789.

Bryan, G. H., and R. Rotunno, 2009b: Evaluation of an Analytical Model for the Maximum Intensity of Tropical Cyclones. J. Atmos. Sci., 66, 3042–3060.

Bryan, G. H., 2012: Effects of Surface Exchange Coefficients and Turbulence Length Scales on the Intensity and Structure of Numerically Simulated Hurricanes. Mon. Wea. Rev., 140, 1125–1143.

Carter, Ashley H. (2001). Classical and Statistical Thermodynamics. Prentice Hall. p. 392. ISBN 0-13-779208-5.

DesRosiers, A. J., Bell, M. M., Klotzbach, P. J., Fischer, M. S., & Reasor, P. D. (2023). Observed relationships between tropical cyclone vortex height, intensity, and intensification rate. Geophysical Research Letters, 50, e2022GL101877.

Dunion, J. P., 2011: Rewriting the climatology of the tropical North Atlantic and Caribbean Sea atmosphere. J. Climate, 24, 893–908.

Drennan, W. M., J. A. Zhang, J. R. French, C. McCormick, and P. G. Black, 2007: Turbulent fluxes in the hurricane boundary layer. Part II: Latent heat flux. J. Atmos. Sci., 64, 1103–1115.

Emanuel, K. A., 1986: An air–sea interaction theory for tropical cyclones. Part I: Steady-state maintenance. J. Atmos. Sci., 43, 585–605.

Emanuel, K. A., 1988: The maximum intensity of hurricanes. J. Atmos. Sci., 43, 1143–1155.

Emanuel, K. A., 1994: Atmospheric Convection. Oxford, 580 pp.

Emanuel, K. A., and R. Rotunno, 2011: Self-stratification of tropical cyclone outflow. Part I: Implications for storm structure. J. Atmos. Sci., 68, 2236–2249.

Emanuel, K., 2012: Self-Stratification of Tropical Cyclone Outflow. Part II: Implications for Storm Intensification. J. Atmos. Sci., 69, 988–996.

Eliassen, A., 1952: Slow Thermally or Frictionally Controlled Meridional Circulation in a Circular Vortex. Astrophys. Norv., 5, 60 pp.

Fairall, C. W., E. F. Bradley, J. E. Hare, A. A. Grachev, and J. B. Edson, 2003: Bulk Parameterization of Air–Sea Fluxes: Updates and Verification for the COARE Algorithm. J. Climate, 16, 571–591.

Fischer, M. S., P. D. Reasor, R. F. Rogers, and J. F. Gamache, 2022: An Analysis of Tropical Cyclone Vortex and Convective Characteristics in Relation to Storm Intensity Using a Novel Airborne Doppler Radar Database. Mon. Wea. Rev., 150, 2255–2278, https://doi.org/10.1175/MWR-D-21-0223.1.

Fischer, M. S., P. D. Reasor, J. P. Dunion, and R. F. Rogers, 2025: Are Rapidly Intensifying Tropical Cyclones Associated with Unique Vortex and Convective Characteristics?. Mon. Wea. Rev., 153, 183–203, https://doi.org/10.1175/MWR-D-24-0118.1.

Hisashi Ozawa & Shinya Shimokawa (2015) Thermodynamics of a tropical cyclone: generation and dissipation of mechanical energy in a self-driven convection system, Tellus A: Dynamic Meteorology and Oceanography, 67:1, 24216, DOI: 10.3402/tellusa.v67.24216

Li, Y., Y. Wang, and Z. Tan, 2024: On the Size Dependence in the Recent Time-Dependent Theory of Tropical Cyclone Intensification. J. Atmos. Sci., 81, 1669–1688.

Montgomery, M. T., and R. K. Smith, 2017: Recent development in the fluid dynamics of the tropical cyclones. Annu. Rev. Fluid Mech., 49, 541–574, https://doi.org/10.1146/annurev-fluid-010816-060022.

Peng, K., R. Rotunno, and G. H. Bryan, 2018: Evaluation of a time dependent model for the intensification of tropical cyclones. J. Atmos. Sci., 75, 2125–2138.

Peng, K., R. Rotunno, G. H. Bryan, and J. Fang, 2019: Evolution of an Axisymmetric Tropical Cyclone before Reaching Slantwise Moist Neutrality. J. Atmos. Sci., 76, 1865–1884.

Persing, J., Montgomery, M. T., McWilliams, J. C., and Smith, R. K., 2013: Asymmetric and axisymmetric dynamics of tropical cyclones, Atmos. Chem. Phys., 13, 12299–12341, https://doi.org/10.5194/acp-13-12299-2013.

Rotunno, R., and K. A. Emanuel, 1987: An air–sea interaction theory for tropical cyclones. Part II: Evolutionary study using a nonhydrostatic axisymmetric numerical model. J. Atmos. Sci., 44, 542–561.

Rotunno, R., 2022: Supergradient Winds in Simulated Tropical Cyclones. J. Atmos. Sci., 79, 2075–2086, https://doi.org/10.1175/JAS-D-21-0306.1.

Schubert, W. H., and J. J. Hack, 1983: Transformed Eliassen Balanced Vortex Model. J. Atmos. Sci., 40, 1571–1583.

Schubert, W. H., S. A. Hausman, M. Garcia, K. V. Ooyama, and H. Kuo, 2001: Potential Vorticity in a Moist Atmosphere. J. Atmos. Sci., 58, 3148–3157, https://doi.org/10.1175/1520-0469(2001)058<3148:PVIAMA>2.0.CO;2.

Smith, R. K., M. T. Montgomery, and S. Vogl, 2008: A critique of Emanuel's hurricane model and potential intensity theory. Quart. J. Roy. Meteor. Soc., 134, 551–561.

Tao, D., R. Rotunno, and M. Bell, 2020: Lilly's Model for Steady-State Tropical Cyclone Intensity and Structure. J. Atmos. Sci., 77, 3701–3720.

Tang, B., and K. Emanuel, 2010: Midlevel ventilation's constraint on tropical cyclone intensity. J. Atmos. Sci., 67, 1817–1830, doi:10.1175/2010jas3318.1.

Tang, B., and K. Emanuel, 2012a: Sensitivity of tropical cyclone intensity to ventilation in an axisymmetric model. J. Atmos. Sci., 69, 2394–2413, doi:10.1175/jas-d-11-0232.1.

Tang, B., and K. Emanuel, 2012b: A ventilation index for tropical cyclones. Bull. Amer. Meteor. Soc., 93, 1901–1912, doi:10.1175/bams-d-11-00165.1.

Wang, D., Y. Lin, and D. R. Chavas, 2022: Tropical Cyclone Potential Size. J. Atmos. Sci., 79, 3001–3025.

Wang, Y., Y. Li, J. Xu, Z. Tan, and Y. Lin, 2021a: The Intensity Dependence of Tropical Cyclone Intensification Rate in a Simplified Energetically Based Dynamical System Model. J. Atmos. Sci., 78, 2033–2045.

Wang, Y., Y. Li, and J. Xu, 2021b: A New Time-Dependent Theory of Tropical Cyclone Intensification. J. Atmos. Sci., 78, 3855–3865.